\documentclass[letterpaper]{JHEP3}
\usepackage{amssymb,amsfonts}
\usepackage{cite}
\usepackage{epsfig}

\newcommand{\ket}[1]{|#1\rangle}
\newcommand{\C}{{\bf C}}
\newcommand{\R}{{\bf R}}

\newcommand{\CA}{{\cal A}}
\newcommand{\CD}{{\cal D}}
\newcommand{\CF}{{\cal F}}
\newcommand{\CG}{{\cal G}}
\newcommand{\CH}{{\cal H}}

\newcommand{\CL}{{\cal L}}
\newcommand{\CM}{{\cal M}}
\newcommand{\CO}{{\cal O}}
\newcommand{\CQ}{{\cal Q}}

\newcommand{\CZ}{{\cal Z}}
\newcommand{\bk}{{\bf k}}
\newcommand{\bx}{{\bf x}}
\newcommand{\by}{{\bf y}}

\newcommand{\FH}{{\mathfrak H}}
\newcommand{\llangle}{\langle\!\langle}
\newcommand{\rrangle}{\rangle\!\rangle}
\newcommand{\LambdaD}{\Lambda_W^{}}
\newcommand{\alpham}{\alpha_{\!M}^{}}

\newcommand{\p}{\partial}
\renewcommand{\bar}[1]{\overline{#1}}
\renewcommand{\tilde}[1]{\widetilde{#1}}

\newcommand{\be}{\begin{equation}}
\newcommand{\ee}{\end{equation}}
\newcommand{\bea}{\begin{eqnarray}}
\newcommand{\eea}{\end{eqnarray}}

\usepackage{graphicx}
\usepackage{latexsym}

\newcommand{\ie}{{\it i.e.}}
\newcommand{\eg}{{\it e.g.}}
\title{Membranes at Quantum Criticality}
\author{Petr Ho\v{r}ava\\
Berkeley Center for Theoretical Physics and Department of Physics\\
University of California, Berkeley, CA, 94720-7300\\
and\\
Theoretical Physics Group, Lawrence Berkeley National Laboratory\\
Berkeley, CA 94720-8162, USA}
\abstract{We propose a quantum theory of membranes designed such that the 
ground-state wavefunction of the membrane with compact spatial topology 
$\Sigma_h$ reproduces the partition function of the bosonic string on 
worldsheet $\Sigma_h$.   The construction involves worldvolume matter at 
quantum criticality, described in the simplest case by Lifshitz scalars with 
dynamical critical exponent $z=2$.  This matter system must be coupled to a 
novel theory of worldvolume gravity, also exhibiting quantum 
criticality with $z=2$.  We first construct such a nonrelativistic 
``gravity at a Lifshitz point'' with $z=2$ in $D+1$ spacetime dimensions, 
and then specialize to the critical case of $D=2$ suitable for the membrane 
worldvolume.  We also show that in the second-quantized framework, the string 
partition function is reproduced if the spacetime ground state takes the form 
of a Bose-Einstein condensate of membranes in their first-quantized ground 
states, correlated across all genera.}

\begin{document}
\section{Introduction}

In the democracy of all branes, strings seem to occupy a privileged position, 
for a variety of reasons.  One argument suggesting that strings are unique 
among all branes points out the apparent nonexistence of the analog of string 
perturbation theory for membranes.  As discussed in \cite{hw1}, this argument 
itself is related to several distinct phenomena.  First, quantum gravity is at 
its critical dimension on the two-dimensional string worldsheet, leading to a 
sensible worldsheet quantum theory at each fixed order in the string 
perturbation expansion.  In contrast, no clear quantization technique 
is available to make sense of quantum gravity coupled to matter on 
higher-dimensional worldvolumes, at least within the conventional approach of 
renormalizable quantum field theory.  Secondly, while two-dimensional 
worldsheets can be organized in terms of a simple discrete invariant -- the 
genus -- which counts the loops of diagrams, no such simple classification is 
available for membranes.  This fact is traditionally interpreted as an 
indication that if a quantum theory of membranes existed, it would have to be 
strongly coupled.  This in turn implies that even if the worldvolume theory on 
a fixed topology were well-defined, we would not know how to sum over distinct 
topologies.  

In this paper, we explore the possibility of constructing a new worldvolume 
quantum theory of gravity and matter in $2+1$ dimensions, at least in the 
simplest case of a bosonic theory.  The price we pay for the avoidance of some 
of the above-mentioned obstacles is a strong anisotropy between space and time 
in the worldvolume theory, a phenomenon familiar from the study of condensed 
matter systems at quantum criticality, dynamical critical phenomena, and in 
statistical dynamics of systems far from equilibrium.   In the process, we 
will uncover a new class of gravity theories with anisotropic scaling between 
space and time, characterized by a nontrivial dynamical critical exponent 
$z$.  Such nonrelativistic gravity models can clearly be of broader interest 
beyond $2+1$ dimensional worldvolumes, and we introduce them first in 
Section~\ref{seclifgrav} in the general case of $D+1$ spacetime dimensions, 
before specializing to $D=2$.  

We begin by posing an auxiliary problem:  
{\it Can we find a quantum theory of membranes, such that its ground-state 
wavefunction reproduces the partition function of the bosonic string?}
This type of question -- about the existence of two systems in such a 
relationship to each other --  is central to many areas of physics, primarily 
with applications to condensed matter.  For example, one might start with a 
universality class describing an equilibrium system in $D$ dimensions at 
criticality, and ask how the critical behavior extends to the dynamical 
phenomena in $D+1$ dimensions.  Requiring that in the static limit one 
recovers the partition function of the original $D$-dimensional equilibrium 
system is effectively equivalent to the type of question that we ask above.  
Essentially the same logic has been used in recent years to produce new 
interesting classes of quantum critical systems in $D+1$ dimensions, starting 
from known classical universality classes in $D$ dimensions.  In stochastic 
quantization, one asks a similar question in imaginary time:  The task is to 
build a nonequilbrium system in $D+1$ dimensions which relaxes at late times 
to its ground state, which reproduces the partition function of the  
$D$-dimensional system one is interested in. The techniques that we use 
in our construction of gravity models are closely related to the methods used 
in these areas of condensed matter theory.  Similar ideas have been applied to 
Yang-Mills gauge theories in \cite{cym}.

Additional motivation for asking our auxiliary question comes also in part 
from the recent findings in topological string theory \cite{witten,gunaydin}, 
the OSV conjecture \cite{osv}, topological M-theory \cite{dijkgraaf}, and 
noncritical M-theory \cite{pixi1,pixi2}.  In that context, interesting 
relationships have been discovered in which the partition function of one 
theory is related to a wavefunction of another theory in a higher dimension, 
and one naturally wonders whether such connections are more prevalent in the 
general context of string and M-theory.     

\section{The Second-Quantized Theory}

In this section, we first define the auxiliary problem a little more 
precisely.  Then, we will assume that the problem is solved at the level of 
first quantization, \ie , that we can construct a membrane worldvolume theory 
whose ground-state wavefunction for the membrane of spatial topology 
$\Sigma_h$ reproduces the partition function of the bosonic string on 
$\Sigma_h$.  Given this assumption, we will show how to solve the problem 
at the second-quantized level, in the Hilbert space of multi-membrane states.  
An attempt to solve the first-quantized problem will then occupy us for the 
rest of the paper.  

In first quantization, our auxiliary question can be interpreted as follows.  
We begin with the critical bosonic string theory in the flat uncompactified 
spacetime with coordinates $X^I$, $I=1,\ldots 26$, described by the Polyakov 
action
\be
W=\frac{1}{4\pi\alpha'}\int d^2\bx\,\sqrt{g}\,g^{ij}\p_iX^I\p_jX^I.
\ee
Its partition function $\CF_h$ on a compact worldsheet $\Sigma_h$ of genus 
$h$ is defined as the path integral 
\be
\label{ppi}
\CF_h=\int_{\CA_h/\CG_h}\CD X(\bx)\,\CD g_{ij}(\bx)\,\exp\left\{-W[X(\bx),
g_{ij}(\bx)]\right\},
\ee
where $\CA_h$ is the space of all fields $X^I(\bx)$ and $g_{ij}(\bx)$ on 
$\Sigma_h$, the gauge group $\CG_h$ consists of worldsheet diffeomorphisms and 
Weyl transformations of $\Sigma_h$, and $\CD X\,\CD g_{ij}$ schematically 
denotes the appropriate measure on the space of gauge orbits $\CA_h/\CG_h$.  
We would like to construct a $2+1$ dimensional quantum theory designed such 
that when it is quantized canonically on $\Sigma_h\times\R$, this theory has 
a ground state whose unnormalized wavefunction $|\Psi_{0h}^{}\rangle$ 
reproduces the string partition function on $\Sigma_h$,
\be
\label{questfirst}
\CF_h=\langle\Psi_{0h}^{}|\Psi_{0h}^{}\rangle.
\ee
More precisely, we will impose a stronger condition which will imply 
(\ref{questfirst}):  Representing the ground state $\ket{\Psi_{0h}^{}}$ in the 
Schr\"odinger representation as a functional 
$\Psi_{0h}^{}[X(\bx),g_{ij}(\bx)]$, we will require that it reproduces 
\be
\label{dens}
\CD X\,\CD g_{ij}\,\exp\left\{-W[X,g_{ij}]\right\}
=\Psi_{0h}^\ast[X,g_{ij}]\Psi_{0h}^{}[X,g_{ij}],
\ee
as an equality between two densities on the space of gauge orbits
$\CA_h/\CG_h$.  The subsequent integral over $\CA_h/\CG_h$ then leads to 
(\ref{questfirst}).  

In the case of $h=0$, the partition function $\CF_0$ of the critical bosonic 
string on $S^2$ vanishes indentically, because the measure in (\ref{ppi}) 
contains the inverse volume of the noncompact conformal Killing symmetry 
group $SL(2,\C)$.  This suggests that any membrane theory which solves 
our first-quantized problem should have no normalizable ground-state 
wavefunction on $S^2\times\R$.

In the rest of this section, we will assume that $\ket{\Psi_{0h}^{}}$ which 
satisfies (\ref{questfirst}) is known, and show that under this assumption, 
the second-quantized problem can be solved by elementary methods of many body 
theory.  We first define the second-quantized string partition function $\CZ$ 
of the closed bosonic string theory to all orders in the string coupling 
$g_s$, 
\be
\label{defz}
\CZ\equiv\exp\left\{\sum_{h=0}^\infty g_s^{2h-2}\CF_h\right\}.
\ee
This expression has a well-defined limit as $g_s\to 0$, because $\CF_0=0$ as 
mentioned above.  

We wish to find a ground state of the second-quantized theory of membranes 
which reproduces this partition function $\CZ$.  This state will be an 
element of the Fock space $\FH$ of multi-membrane states.  (We denote states 
in this second-quantized Hilbert space by $|\ \rrangle$, to distinguish 
them from the first-quantized quantum states of a single membrane.)  Thus, 
we are looking for a ground state $|\Psi_0\rrangle$ which satisfies
\be
\label{questsecond}
\CZ=\llangle\Psi_0|\Psi_0\rrangle.
\ee 

We choose to present the solution to the second-quantized problem first, 
because the answer is robust and rather insensitive to the precise form of the 
solution to the first-quantized problem.  In fact, it is natural to expect 
that if the first-quantized problem has a solution, it will not be unique -- 
two distinct theories on the membrane worldvolume can share the same ground 
state but differ in their spectra of excited states.  This is analogous to the 
relationship between static and dynamical critical phenomena:  A single static 
universality class can split into several distinct dynamical universality 
classes, which all share the same equilibrium properties.  

\subsection{The Fock Space of Membranes}

Imagine that we have been given a basis of single-membrane quantum states, 
\be
\ket{\Psi_{0h}^{}},\qquad \left\{\ket{\Psi_{\alpha h}^{}}\right\},
\ee
where the index $\alpha$ denotes collectively all the excited states, 
and the ground state $\ket{\Psi_{0h}^{}}$ satisfies (\ref{questfirst}).
The second-quantized Hilbert space $\FH$ of multi-membrane states can then be 
constructed by elementrary methods of many-body physics.  First we associate 
a pair of creation and annihilation operators with each state,
\be
A_{0h}^{},A^\dagger_{0h},\qquad A_{\alpha h}^{},A^\dagger_{\alpha h}.
\ee
These satisfy the canonical commutation relations,
\be
[A_{0h}^{},A^\dagger_{0h'}]=\delta_{hh'},\qquad [A_{\alpha h}^{},
A^\dagger_{\beta h'}]=\delta_{\alpha\beta}\delta_{hh'},
\ee
(with all the unlisted commutators equal to zero), and can be used to define 
the second-quantized Fock space $\FH_h$ of quantum membranes of 
genus $h$, by first defining the Fock space vacuum $|0\rrangle_{\!h}$ via
\be
A_{0h}^{}|0\rrangle_{\!h}=0,\qquad A_{\alpha h}^{}|0\rrangle_{\!h}=0. 
\ee
The Fock space $\FH_h$ is then built in the standard way by the action of 
the creation operators $A_{0h}^\dagger$, $A_{\alpha h}^\dagger$ on 
$|0\rrangle_{\!h}$.  Each creation operator creates a membrane in the 
corresponding quantum state $\ket{\Psi_{0h}^{}}$ or $\ket{\Psi_{\alpha h}^{}}
$.  States in $\FH_h$ thus correspond to collections of an arbitrary number 
of quantum membranes, each of genus $h$.  

The total Hilbert space $\FH$ is the tensor product of $\FH_h$ 
over all values of $h$,
\be
\FH\equiv\bigotimes_{h=0}^\infty\FH_h,
\ee
The total Hilbert space $\FH$ is the Fock space generated by the application 
of arbitrary collections of creation operators on the Fock vacuum, defined as
\be
|0\rrangle\equiv\bigotimes_{h=0}^\infty|0\rrangle_{\!h}.
\ee

\subsection{Bose-Einstein Condensation and Spacetime Superfluidity}

The second-quantized Fock space $\FH$ contains multi-membrane 
states, with any number of membranes of any genera.  We claim that our 
desired state $|\Psi_0\rrangle$ is a specific state in $\FH$, given by
\be
\label{grstate}
|\Psi_0\rrangle=\bigotimes_{h=0}^\infty
\exp\left\{g_s^{h-1}A^\dagger_{0h}\right\}|0\rrangle_{\!h}=
\exp\left\{\sum_h g_s^{h-1}A^\dagger_{0h}\right\}|0\rrangle.
\ee
A direct calculation using (\ref{questfirst}) and the elementary algebra of 
creation and annihilation operators shows that (\ref{grstate}) indeed 
satisfies the desired property (\ref{questsecond}).    

This state has an interesting intuitive interpretation.  For each value of 
$h$, this state looks like the ground state of a spacetime theory in which 
membranes of genus $h$ -- all in their ground state $\ket{\Psi_{0h}^{}}$ -- 
have formed a spacetime Bose-Einstein condensate.  Defining the number 
operators in each membrane sector via
\be
N_{0h}=A_{0h}^\dagger A_{0h}^{},\qquad
N_{\alpha h}=A_{\alpha h}^\dagger A_{\alpha h}^{},
\ee
we see that the proposed ground state (\ref{grstate}) is not an eigenstate 
of $N_{0h}$ and thus does not contain a definite finite number of membranes. 

Instead, the ground state is an eigenstate of $A_{0h}^{}$, with eigenvalue 
$g_s^{h-1}$.  This indicates that the strength of the condensate of membranes 
of different genera is correlated over all $h$, via the value of the string 
coupling constant $g_s$.  It would be interesting to study model Hamiltonians 
that reproduce the same ground state.  Such an analysis would be difficult in 
the absence of at least some information about the membrane excited states.  
However, the correlated nature of the condensate suggests that in this 
second-quantized ground state, membranes interact via a local contact 
interaction, which is insensitive to the global topology of the membranes.  
This interaction allows processes in which a membrane of genus $h$ gets 
pinched into a pair of genera $h'$ and $h-h'$ respectively, or the 
self-pinching in which the genus changes to $h-1$, plus the reversal of these 
two processes.  

In conventional condensed matter systems, ground states that take the form of 
a Bose-Einstein condensate typically lead to superfluidity, characterized by 
gapless excitations with an emergent relativistic low-energy dispersion 
relation.  It is tempting to predict that the membrane theory whose ground 
state is given by (\ref{grstate}) similarly exhibits spacetime superfluidity.  
However, the knowledge of the ground state (\ref{grstate}) itself is not 
sufficient to determine whether the system has excitations that behave as 
those of a superfluid.  As pointed out above, it is possible that 
different solutions of the first-quantized problem might exist, sharing the 
same single-membrane ground state $\ket{\Psi_{0h}^{}}$ but with different 
spectra of excited states $\ket{\Psi_{\alpha h}^{}}$.  For example, a minimal 
solution to our first-quantized problem would be given by a theory of 
membranes where the ground state $\ket{\Psi_{0h}^{}}$ is the only physical 
state on $\Sigma_h$, and the membrane has no physical excited states (and 
in particular, no states carrying nonzero values of spatial momenta).  
In such a minimal realization, the spacetime theory would be effectively a 
topological theory, and the Bose-Einstein condensate would not be accompanied 
by spacetime superfluidity.  

\section{The First-Quantized Theory}

Having shown how the second-quantized problem is solved assuming the existence 
of a worldsvolume theory that reproduces the worldsheet path integral of 
string theory genus by genus, it now remains to solve the corresponding 
first-quantized problem.  

\subsection{Worldvolume Matter: Lifshitz Scalars and Quantum Criticality} 
\label{seclifs}

We will pose the question first for a single worldsheet scalar field, before 
coupling to worldsheet gravity.  In fact, it will be useful to take the 
broader perspective and consider a scalar field theory in $D$ flat Euclidean 
dimensions $\bx=(x^i)$, $i=1,\ldots D$, with the Euclidean action   
\be
\label{euscalar}
W=\frac{1}{2}\int d^D\bx\,\left(\p_i\Phi\p_i\Phi\right).
\ee
The partition function of the free scalar takes the form 
\be
\label{scpth}
\CZ=\int\CD\Phi(\bx)\exp\left\{-W[\Phi(\bx)]\right\},  
\ee
of a path integral on the space of field configurations $\Phi(\bx)$.  
Imagine now a theory in $D+1$ dimensions whose configuration space coincides 
with the space of all $\Phi(\bx)$.  In the Schr\"odinger representation, the 
wavefunctions of this theory are functionals $\Psi[\Phi(\bx)]$.  For any given 
wavefunction, $|\Psi[\Phi(\bx)]|^2$ is naturally a density on the 
configuration space.  We want to design our system such that its ground-state 
wavefunction $\Psi_0[\Phi(\bx)]$ reproduces the path integral density of 
(\ref{scpth}), 
\be
\label{square}
\CD\Phi(\bx)\exp\left\{-W[\Phi(\bx)]\right\}=\Psi_0^\ast[\Phi(\bx)]\,
\Psi_0[\Phi(\bx)].
\ee

As it turns out, a construction which yields the desired answer for the 
scalar field is known in the condensed matter literature (see, \eg , 
\cite{ardonne}).  It is given in terms of a slightly exotic scalar field 
theory in $D+1$ dimensions, whose action is
\be
\label{lifscalaract}
S=\frac{1}{2}\int dt\,d^D\bx\left\{(\dot\Phi)^2-\frac{1}{4}(\Delta
\Phi)^2\right\}.
\ee
Here $\Delta$ is the spatial Laplacian, $\Delta=\p_i\p_i$.  Note that $S$ is 
a sum of a ``kinetic term'' involving time derivatives, and a ``potential 
term'' which is of a special form:  It can be derived from 
a variational principle, 
\be
\frac{1}{4}(\Delta\Phi(\bx))^2=
\left(\frac{1}{2}\frac{\delta W}{\delta\Phi(\bx)}\right)^2,
\ee
where $W$ is the action (\ref{euscalar}) of the Euclidean scalar theory.
Henceforth we say that theories that enjoy this property satisfy the 
``detailed balance'' condition.  This property, and its extension to the case 
involving gravity, will play a central role in the rest of the paper.

The scalar field theory (\ref{lifscalaract}) is a prototype of a class of 
models introduced and studied in the context of tri-critical 
phenomena in condensed matter physics by Lifshitz \cite{lifshitz} in 1941, and 
is consequently referred to in the literature as the ``Lifshitz scalar'' field 
theory \cite{hornreich,lubensky}.  In the context originally studied by 
Lifshitz \cite{lifshitz}, $t$ is Wick rotated to become one 
of the spatial dimensions, and the Lifshitz scalar then describes the 
tricritical point connecting the phases with a zero, homogeneous or 
spatially modulated condensate of $\Phi$.  The same theory is also relevant 
in the description of various universality classes in dynamical critical 
phenomena, and in quantum criticality.  In particular, the Lifshitz scalar 
is believed to be in the same universality class as the quantum dimer problem, 
which is particularly intriguing because of the close connection 
\cite{okounkov,pixi2} between dimer models, topological string theory and 
noncritical M-theory.  

Why is the ground-state wavefuction of the Lifshitz scalar theory related to 
the partition function (\ref{scpth})?  The key to this fact is the detailed 
balance condition obeyed by the Lifshitz scalar.  To identify the ground-state 
wavefunction, we quantize the theory canonically.  The Hamiltonian of the 
Lifshitz scalar is
\be
H=\frac{1}{2}\int d^D\bx\left\{P^2+\frac{1}{4}(\Delta\Phi)^2\right\}.  
\ee
In the Schr\"odinger representation we realize the momenta conjugate to 
$\Phi(\bx)$ as
\be
P(\bx)=-i\frac{\delta}{\delta\Phi(\bx)}.
\ee
Up to a normal-ordering constant, the Hamiltonian can be written as
\be
\label{haml}
H=\frac{1}{2}\int d^D\bx\left(-\frac{\delta}{\delta\Phi}-\frac{1}{2}
\Delta\Phi\right)\left(\frac{\delta}{\delta\Phi}-\frac{1}{2}\Delta\Phi\right)
=\int d^D\bx\,\bar\CQ\CQ,
\ee
where
\be
\label{cqvar}
\CQ(\bx)=iP(\bx)-\frac{1}{2}\Delta\Phi(\bx)=
\frac{\delta}{\delta\Phi(\bx)}-\frac{1}{2}\Delta\Phi(\bx),
\ee
and $\bar\CQ$ is its complex conjugate.  Consequently, any functional 
$\Psi_0[\Phi(\bx)]$ that is annihilated by $\CQ$,
\be
\label{firstq}
\CQ\Psi_0[\Phi(\bx)]\equiv\left(\frac{\delta}{\delta\Phi(\bx)}
+\frac{1}{2}\Delta\Phi(\bx)\right)\Psi_0[\Phi(\bx)]=0,
\ee
is an eigenstate of the Hamiltonian with the lowest eigenvalue, and thus 
represents a candidate wavefunction of the ground state. In order for this 
candidate to be a true wavefunction, it must be normalizable.  

Because the Lifshitz scalar theory satisfies the detailed balance condition, 
it is easy to find a simple solution to (\ref{firstq}), given by
\be
\Psi_0[\Phi(\bx)]=\exp\left\{-\frac{1}{4}\int d^D\bx\,\p_i\Phi\p_i\Phi\right\}.
\ee
This is a normalizable wavefunction of the ground state, which in turn yields 
(\ref{square}) and solves our problem.  

The fact that the Hamiltonian (\ref{haml}) can be written as 
$\int \bar\CQ\CQ$, and the subsequent role played by the simpler condition 
$\CQ\Psi_0=0$ in indentifying the lowest energy eigenstates, are reminiscent 
of supersymmetry and the role played by the BPS condition.  This resemblance 
is not accidental, and can be rephrased in terms of an underlying 
supersymmetry with scalar supercharges, formally similar to topological BRST 
symmetry. In the context of condensed matter applications mentioned above, 
this symmetry is known as the Parisi-Sourlas supersymmetry \cite{ps} (see 
also \cite{sourlas} for a nice early review).  In the context of strings and 
membranes, Parisi-Sourlas supersymmetry played a role in \cite{rigid}.  We 
will not use the supersymmetric formalism in the present paper.  

Regardless of its relation with the Euclidean scalar theory in $D$ dimensions, 
the Lifshitz scalar theory in $D+1$ dimensions is an interesting system in its 
own right.  Its action (\ref{lifscalaract})  defines a Gaussian RG fixed 
point, with scaling properties which are somewhat exotic from the perspective 
of relativistic quantum field theory.  We will measure the scaling properties 
of various quantities in the units of inverse spatial length.  In order for 
the two terms in the action to scale the same way, we must assign anisotropic 
scaling properties to space and time,  
\be
[\bx]=-1,\qquad [t]=-2.
\ee
In condensed matter systems, the degree of anisotropy between time and space 
is measured by the dynamical cricial exponent $z$.  Lorentz symmetry in 
relativistic systems implies $z=1$, while nonrelativistic systems with 
Galilean invariance have $z=2$.  More generally, the dynamical critical 
exponent can be defined in terms of the scaling properties of two-point 
functions,
\be
\langle\Phi(\bx,t)\Phi(0,0)\rangle=\frac{1}{|\bx|_{}^{2[\Phi]}}\,f\left(
\frac{\bx}{t^{1/z}}\right),
\ee
where $[\Phi]$ is the conformal dimension of $\Phi$.  In the case of the free 
Lifshitz scalar theory, we have $z=2$, and 
\be
[\Phi]=\frac{D-2}{2}.
\ee
This conformal dimension is of course different from the dimension 
$[\Phi]_{z=1}$ of the scalar field at the relativistic Gaussian fixed point 
in $D+1$ dimensions, which is $[\Phi]_{z=1}=(D-1)/2$.  As a result, the lower 
critical dimension of the Lifshitz scalar at which the two-point function 
becomes logarithmic is $2+1$, and not $1+1$ as in the usual relativistic 
case.%
\footnote{Straightforward generalizations of the Lifshitz theory exist 
\cite{lubensky}, such as theories at the $(m,n)$ Lifshitz point, with 
$m$ dimensions like $t$ and $n$ dimensions like $\bx$.}
Remarkably, making the system anisotropic causes a shift in the critical 
dimension of the system.  

In the case of relativistic scalar field theory, the importance of $1+1$ being 
the critical dimension can hardly be overstated.  This fact is at the core of 
string theory, and represents perhaps the most elegant way 
\cite{friedan1,friedan2} of deriving Einstein's equations and their systematic 
higher-order corrections, from the simple condition of quantum conformal 
invariance of the nonlinear sigma model.  Similarly, one can can generalize 
the Lifshitz scalar theory to an anisotropic nonlinear sigma model, which will 
have an infinite number of classically marginal couplings in $2+1$ 
dimensions.  A detailed study of the RG properties of such Lifshitz-type sigma 
models should be very interesting.

\subsection{Requirements on Worldvolume Gravity} 

In order to extend the construction from the matter sector to the full 
string worldsheet theory, we need to couple the Lifshitz scalar theory to 
some form of worldvolume gravity.  When this worldvolume system is quantized 
on $\Sigma_h\times\R$, the resulting wavefunction of the membrane ground state 
is supposed to reproduce (\ref{dens}).  Consequently, the ground-state 
wavefunction must be a functional of $X^I(\bx)$ and $g_{ij}(\bx)$ defined on 
the space of gauge orbits $\CA_h/\CG_h$:  In other words, $\Psi_0$ must be 
invariant under worldsheet diffeomorphisms and Weyl transformations.    

Our task is to design a gravity theory in $2+1$ dimensions which reproduces 
these expected properties of the ground-state wavefunction, much like the 
Lifshitz scalar reproduces the path integral of the worldsheet matter sector.  
This gravity theory should naturally couple to the anisotropic theory of 
matter described by Lifshitz scalars with $z=2$.  In order to match the 
scaling properties of the matter sector, this gravity theory should therfore 
also be at quantum criticality with $z=2$.  

The possibility of constructing a nonrelativistic theory of gravity with 
anisotropic scaling and nontrivial values of $z$ is clearly of a more general 
interest.  Therefore, we devote Section~\ref{seclifgrav} to the presentation 
of such anisotropic gravity models in the general case of $D+1$ dimensions, 
and return to $D=2$ in Section~\ref{secmemc}.  
\vfill\break

\section{Gravity at a $z=2$ Lifshitz Point in $D+1$ Dimensions}
\label{seclifgrav}

In this section, we formulate a classical theory of gravity with dynamical 
critical exponent $z=2$.    As in the case of the Lifshitz scalar reviewed 
in Section~\ref{seclifs}, it will be instructive to consider our construction 
in $D+1$ dimensions, specializing to the case of $D=2$ only later as required 
for the application to the membrane worldvolume.   

We will assume that our spacetime is topologically of the form 
$\CM=\R\times\Sigma$, where $\Sigma$ is a compact $D$-dimensional space.  This 
assumption will simplify our construction, by avoiding the discussion of 
the possible spatial boundary terms in the action.  

\subsection{First Ingredients}

As a minimal requirement, our theory in $D+1$ dimensions should describe 
spatial components $g_{ij}(\bx,t)$ of the metric, $i,j=1,\ldots D$.  The gauge 
symmetries will surely have to contain diffeomorphisms of space.  Motivated 
by the form of the Lagrangian for the Lifshitz scalar, our gravity theory will 
have a kinetic term given by
\be
\label{kin}
S_K=\frac{1}{2\kappa^2}\int dt\,d^D\bx\sqrt{g}\,\dot g_{ij}\,G^{ijk\ell}\,
\dot g_{k\ell}.
\ee
We have introduced a coupling constant $\kappa$, whose physical role will 
become clear later, in Section~\ref{secffpt}.  Throughout the paper, we use 
``$\dot{\ \ }$'' to denote the time derivative; 
\eg , $\p_t g_{ij}\equiv\dot g_{ij}$.  

In order to write down this kinetic term, we needed a ``metric on the space of 
metrics,'' denoted here by $G^{ijk\ell}$.  Spatial diffeomorphism 
invariance of the action requires $G^{ijk\ell}$ to take, up to an overall 
normalization, the following form  
\be
\label{dwtt}
G^{ijk\ell}=\frac{1}{2}\left(g^{ik}g^{j\ell}+g^{i\ell}g^{jk}\right)
-\lambda\, g^{ij}g^{k\ell},
\ee
with $\lambda$ an arbitrary real constant.  This object is very similar to 
the familiar De~Witt metric of general relativity.  In the relativistic 
theory, the full spacetime diffeomorphism invariance fixes the value of 
$\lambda$ uniquely, to equal $\lambda=1$.  In that case, the ``metric on the 
space of metrics'' $G^{ijk\ell}$ is known as the ``De~Witt metric.''  We will 
extend this terminology to our more general case as well, even when 
$\lambda$ is not necessarily equal to one.  

For now, $\lambda$ plays the role of a coupling constant.  In 
Section~\ref{secmemc} we will see that the value of this coupling is uniquely 
determined, if we require that the theory also respect an anisotropic version 
of local Weyl invariance.  

\subsection{The Potential Term}

In our gravity theory, we wish to maintain the anisotropy of scaling between 
space and time, consistent with the value of dynamical critical exponent $z=2$ 
of the Lifshitz matter fields.  As a result, we look for a ``potential term'' 
$S_V$ of fourth-order in spatial derivatives, so that the full action of our 
system is the sum of two parts,
\be
S=S_K-S_V.
\ee  
In principle, if we follow the usual logic of effective field theory, many 
such terms can be written down and therefore should be included in the 
effective action.  However, the choices can be severely reduced, if we require 
the existence of an action $W[g_{ij}(\bx)]$ in $D$ dimensions such that
\be
\label{pot}
S_V=\frac{\kappa^2}{2}\int dt\,d^D\bx\,\sqrt{g}\,E^{ij}\,\CG_{ijk\ell}\,
E^{k\ell},
\ee
where $E^{ij}$ follows from the variational principle with action 
$W[g_{ij}(\bx)]$,
\be
\label{princ}
\sqrt{g}\,E^{ij}=\frac{1}{2}\frac{\delta W}{\delta g_{ij}}.
\ee
In other words, we require that our potential term satisfies the gravitational 
analog of the ``detailed balance'' condition mentioned in 
Section~\ref{seclifs}.  

In (\ref{pot}), $\CG_{ijk\ell}$ denotes the inverse of the De~Witt metric, 
\be
\CG_{ijmn}G^{mnk\ell}=\frac{1}{2}\left(\delta_i^k\delta_j^\ell+
\delta_i^\ell\delta_j^k\right).  
\ee
More explicitly, we have
\be
\CG_{ijk\ell}=\frac{1}{2}\left(g_{ik}g_{j\ell}+g_{i\ell}g_{jk}
\right)-\tilde\lambda\,g_{ij}g_{k\ell},
\ee
with 
\be
\tilde\lambda=\frac{\lambda}{D\lambda-1}.
\ee

In order to end up with $S_V$ which is invariant under spatial diffeomorphisms 
and of fourth order in spatial derivatives, we must take $W$ to be the 
Einstein-Hilbert action in $D$ dimensions,%
\footnote{Our notation in this paper is strictly nonrelativistic:  All 
quantities such as the covariant derivative $\nabla_i$, the Ricci scalar $R$,  
the Ricci tensor $R_{ij}$ {\it etc.}, are always defined in terms of the 
metric $g_{ij}$ on the $D$-dimensional leaves of the spacetime foliation, 
unless explicitly stated otherwise.}%
\be
\label{ehact}
W=\frac{1}{\kappa_W^2}\int d^D\bx\,\sqrt{g}\,R.
\ee
The general action $W$ could also contain a cosmological constant term 
$\LambdaD$.  However, for now, we set $\LambdaD=0$ in order to focus on the 
leading term in $W$ that produces the dominant, highest-dimension operators in 
$S_V$.  We will return to the discussion of the general case with nonzero 
$\LambdaD$ in Section~\ref{seclowd}.  

With the choice of the Einstein-Hilbert term (\ref{ehact})as $W$, the full 
action is given by
\bea
\label{redact}
S&=&\frac{1}{2}\int dt\,d^D\bx\,\sqrt{g}\left\{\frac{1}{\kappa^2}\dot g_{ij}
G^{ijk\ell}\dot g_{k\ell}-\frac{\kappa^2}{4\kappa_W^4}\left(R^{ij}
-\frac{1}{2}Rg^{ij}\right)\CG_{ijk\ell}\left(R^{k\ell}-\frac{1}{2}Rg^{k\ell}
\right)\right\}\cr
&&\qquad{}=\frac{1}{2}\int dt\,d^D\bx\,\sqrt{g}\left\{\frac{1}{\kappa^2}
\dot g_{ij}G^{ijk\ell}\dot g_{k\ell}-\frac{\kappa^2}{4\kappa_W^4}\left(
R^{ij}R_{ij}+aR^2\right)\right\},
\eea
where $a$ is a constant equal to
\be
\label{alpha}
a=\frac{1-\lambda-D/4}{D\lambda-1}.
\ee
Note that in a large range of values of $D$ and $\lambda$ the potential term 
$S_V$ in the action is manifestly positive definite.  

\subsection{Extending the Gauge Symmetries}

The action in (\ref{redact}) appears to be a good first step, but it is only 
invariant under {\it spatial\/} diffeomorphisms  
\be
\delta x^i=\zeta^i(x^j)
\ee
and global time translations, and there is no Weyl invariance.  As a result, 
when we specialize to $D=2$, the hypothetical ground-state wavefunction would 
depend on the conformal factor of the two-dimensional metric.  Thus, we will 
need to accomodate Weyl invariance in order to make contact with the partition 
function of critical string theory.  The way to resolve these issues is to 
require extended gauge symmetries, which will in turn require new gauge 
fields. 

\subsubsection{Foliation-Preserving Diffeomorphisms}

Given the preferred role of time in our theory, it is natural to extend 
the gauge symmetry of time-independent spatial diffeomorphisms enjoyed by 
(\ref{redact}) to all spacetime diffeomorphisms that respect the preferred 
codimension-one foliation%
\footnote{A codimension-$q$ foliation $\CF$ of a $d$-dimensional manifold 
$\CM$ is defined as $\CM$ equipped with an atlas of coordinate systems 
$(y^a,x^i)$ $a=1,\ldots q$, $i=1,\ldots d-q$, such that the transition 
functions take the restricted form $(\tilde y^a,\tilde x^i)=
(\tilde y^a(y^b),\tilde x^i(y^b,x^j))$.  For the general theory of 
foliations, see \eg\ \cite{lawson,godbillon,moerdijk} and references therein.}
$\CF$ of spacetime $\CM$ by the slices of fixed time.  Such 
``foliation-preserving diffeomorphisms'' will consist of spacetime-dependent 
spatial diffeomorphisms as well as time-dependent time reparametrizations, 
generated by infinitesimal transformations
\be
\label{fdif}
\delta x^i=\zeta^i(t,\bx),\qquad \delta t=f(t).
\ee

Together with the new symmetries, we also introduce new fields, $N$ and 
$N_i$.  From the point of view of the $D+1$ canonical ADM formalism in 
relativistic gravity, these are the well-known ``lapse and shift'' variables.  
Thus, our theory will share its field content with conventional relativistic 
gravity theory, at least if $N$ and $N_i$ are allowed to be functions of both 
space and time.   

\subsubsection{Spacetime Diffeomorphisms and the Nonrelativistic Limit}

The algebra of foliation-preserving diffeomorphisms and its action on the 
fields $g_{ij}$, $N_i$ and $N$ can be conveniently derived from the 
relativistic action of all diffeomorphisms on $g_{\mu\nu}$, by restoring the 
speed of light $c$ and taking the nonrelativistic limit $c\to\infty$.  We 
start with the relativistic metric $g_{\mu\nu}$ in the usual ADM decomposition 
but with $c$ restored,
\be
g_{\mu\nu}=\pmatrix{-N^2+N_iN^i/c^2,& N_i/c\cr
N_i/c,& g_{ij}},
\ee
and with $x^0=ct$.  One can view this expression as a leading order of an 
expansion in $1/c$.  Similarly, we expand the generators $v^\mu$ of spacetime 
diffeomorphisms in the powers of $1/c$,
\be
\label{diffexp}
v^t=cf(t,\bx)+\CO(1/c),\qquad v^i=\zeta^i(t,\bx)+\CO(1/c^2).
\ee
In order to obtain a nonsingular $c\to\infty$ limit, the generator of time 
reparametrizations $f$ in (\ref{diffexp}) must be restricted to be a function 
of $t$ only.   With this condition, the standard action of relativistic 
diffeomorphism generators $v^\mu$ on $g_{\mu\nu}$ contracts to the 
diffeomorphisms (\ref{fdif}) that preserve the preferred foliation of 
spacetime by leaves of constant time $t$.  Their action on the component 
fields is obtained by taking the $c\to\infty$ limit of the relativistic 
diffeomorphisms $v^\mu$ acting on $g_{\mu\nu}$, which leads to 
\bea
\label{foldif}
\delta g_{ij}&=&\p_i\zeta^kg_{jk}+\p_j\zeta^kg_{ik}+\zeta^k\p_kg_{ij}+f\,
\dot g_{ij},\nonumber\\
\delta N_i&=&\p_i\zeta^jN_j+\zeta^j\p_jN_i+\dot\zeta^jg_{ij}+\dot f\,N_i+f
\,\dot N_i,\\
\delta N&=&\zeta^j\p_jN+\dot f\,N+f\,\dot N.\nonumber
\eea
In order to obtain a smooth $c\to\infty$ limit, $f$ can only be a function of 
$t$, while $\zeta^i$ is allowed to depend on both $t$ and $x^j$:  The algebra 
becomes that of the foliation-preserving diffeomorphisms (\ref{fdif}).  Note 
that the transformation rules under foliation-preserving diffeomorphisms 
do not depend of the anticipated value of the dynamical exponent $z$ that 
measures the degree of anisotropy between space and time.  Thus, the value of 
$z$ is not determined by the gauge symmetries, and represents an interesting 
dynamical quantity in our theory.  

Because the generator of time diffeomophisms $f(t)$ is a function of time 
only, the gauge symmetry of foliation-preserving diffeomorphisms has one less 
generator per spacetime point than general diffeomorphism symmetry.  It is 
natural to match this by restricting the corresponding gauge field $N$, 
associated with the time diffeomorphisms, to also be a function of only $t$.  
This step is not strictly mandated by the structure of the symmetry 
transformations (\ref{foldif}), but allowing $N$ to be a general function 
of $t$ and $\bx$ would lead to difficulties in quantization, at least in the 
absence of extra gauge symmetries.

There is an interesting possibility of taking the nonrelativistic limit in 
such a way that the number of local symmetries matches that of general 
relativity.  It involves keeping the subleading term in the $1/c$ expansion 
of the time-time component of the metric,%
\footnote{This is the place which would be occupied in the nonrelativistic 
expansion of general relativity by the Newton potential.}  
$g_{00}=-N^2+(N_iN^i+2A)/c^2$, and keeping the subleading term in the time 
component of the diffeomorphism transformation, 
$v^t=cf(t)-\varepsilon(t,\bx)/c$.  It turns out that $\varepsilon(t,\bx)$ 
acts on the fields by
\bea
\label{uone}
\delta_\varepsilon A&=&N^2\dot\varepsilon+N\dot N\varepsilon-N^2N^i\p_i
\varepsilon,\nonumber\\
\delta_\varepsilon N_i&=&N^2\p_i\varepsilon,\\
\delta_\varepsilon N&=&\delta_\varepsilon g_{ij}=0.\nonumber
\eea
If the leading term $N(t)$ in $g_{00}$ is restricted to be only a function of 
time as suggested above, this new symmetry is simply an Abelian gauge symmetry 
with gauge parameter $N\varepsilon$, and with $A/N$ and $N_i/N$ transforming 
as an Abelian connection.  However, extending the foliation-preserving 
diffeomorphisms by this Abelian gauge symmetry appears to run into 
difficulties with constructing nontrivial Lagrangians invariant under this 
symmetry, and we will not pursue the possibility of such an extended gauge 
symmetry in this paper.  

\subsubsection{The Covariant Action}

In order to make our theory invariant under foliation-preserving 
diffeomorphisms, we need to decorate various terms in the action 
(\ref{redact}) by the appropriate dependence on $N$ and $N_i$.  For example, 
the covariant volume element is $\sqrt{g}N$, and the time derivative of the 
metric is replaced by 
\be
\dot g_{ij}\rightarrow \frac{1}{N}\left(\dot g_{ij}-\nabla_iN_j-\nabla_j N_i
\right),
\ee
which transforms covariantly under foliation-preserving diffeomorphisms.
Similarly, any term of the form
\be
\int dt\,d^D\bx\,\sqrt{g}\,V[g_{ij}]
\ee
which respects time-independent spatial diffeomorphisms and does not depend 
on time derivatives of $g_{ij}$ can be covariantized as
\be
\int dt\,d^D\bx\,\sqrt{g}\,N\,V[g_{ij}].
\ee
Indeed, we have
\be
\delta_f\int dt\,d^D\bx\,\sqrt{g}\,N\,V[g_{ij}]=\int dt\,d^D\bx\,
\p_t\left(\sqrt{g}\,f\,N\,V[g_{ij}]\right).
\ee
In the end, the full covariant action is 
\bea
\label{fullact}
S&=&\frac{1}{2}\int dt\,d^D\bx\,\sqrt{g}\left\{\frac{1}{\kappa^2N}
\left(\dot g_{ij}-\nabla_iN_j-\nabla_jN_i\right)G^{ijk\ell}
\left(\dot g_{k\ell}-\nabla_kN_\ell-\nabla_\ell N_k\right)
\right.\cr
&&\qquad\qquad\qquad\left.{}-\frac{\kappa^2N}{4\kappa_W^4}\left(R^{ij}
-\frac{1}{2}Rg^{ij}\right)\CG_{ijk\ell}\left(R^{k\ell}-\frac{1}{2}Rg^{k\ell}
\right)\right\}.
\eea
Setting $N=1$, $N_i=0$ would restore the reduced action (\ref{redact}).

\subsubsection{Detailed Balance Condition}
\label{secdetbal}

As is the case for relativistic quantum field theories, explicit calculations 
are most conveniently performed after the Wick rotation to imaginary time, 
$\tau=it$.  This rotation entails $N_j\rightarrow iN_j$.  After the rotation, 
the action can be rewritten -- up to total derivatives -- as a sum of squares,%
\footnote{In the Wick-rotated theory, ``$\dot{\ \ }$'' denotes differentiation 
with respect to the imaginary time $\tau$.} 
\bea
\label{imact}
S&=&\frac{i}{2}\int d\tau\,d^D\bx\,\sqrt{g}N\left\{\left[\frac{1}{\kappa N}
\left(\dot g_{ij}-\nabla_iN_j-\nabla_jN_i\right)+\frac{\kappa}{2\kappa_W^2}
\CG_{ijmn}\left(R^{mn}-\frac{1}{2}Rg^{mn}\right)\right]\right.\cr
&&\qquad\left.{}\times G^{ijk\ell}
\left[\frac{1}{\kappa N}\left(\dot g_{k\ell}-\nabla_kN_\ell-\nabla_\ell 
N_k\right)+\frac{\kappa}{2\kappa_W^2}\CG_{k\ell pq}\left(R^{pq}
-\frac{1}{2}Rg^{pq}\right)\right]\right\}.
\eea
In order to see that (\ref{fullact}) is indeed reproduced from (\ref{imact}) 
by the inverse Wick rotation, we need to show that the cross-terms
\be
\int d\tau\,d^D\bx\,\sqrt{g}N\left\{\left[\frac{1}{\kappa N}
\left(\dot g_{ij}-\nabla_iN_j-\nabla_jN_i\right)\right]G^{ijk\ell}
\left[\frac{\kappa}{2\kappa_W^2}\CG_{k\ell pq}\left(R^{pq}
-\frac{1}{2}Rg^{pq}\right)\right]\right\}
\ee
are a sum of total derivatives.  First, we have
\bea
&&\int d\tau\,d^D\bx\,\sqrt{g}N\left\{\frac{1}{\kappa N}
\dot g_{ij}G^{ijk\ell}\left[\frac{\kappa}{2\kappa_W^2}\CG_{k\ell pq}
\left(R^{pq}-\frac{1}{2}Rg^{pq}\right)\right]\right\}\cr
&&\qquad{}=-\frac{1}{2}\int d\tau\,d^D\bx\,\dot g_{ij}
\frac{\delta W}{\delta g_{ij}}=-\frac{1}{2}\int d\tau\,d^D\bx\,\p_\tau
(\CL_W^{}),
\eea
where $\CL_W^{}$ is the Lagrangian density, $W=\int d\tau\,d^D\bx\,\CL_W^{}$.  
For this to hold, it was crucial that (i) the potential term $S_V$ is a 
square of terms (\ref{princ}) which originate from a variational principle, 
and (ii) that the metric $\CG_{ijk\ell}$ used in the potential term $S_V$ is 
the inverse of the De~Witt metric $G^{ijk\ell}$ that appeared in the kinetic 
term $S_K$.  

Similarly, 
\bea
&&\int d\tau\,d^D\bx\,\sqrt{g}N\left[\frac{1}{\kappa N}
\left(\nabla_iN_j+\nabla_jN_i\right)\right]G^{ijk\ell}
\left[\frac{\kappa}{2\kappa_W^2}\CG_{k\ell pq}\left(R^{pq}
-\frac{1}{2}Rg^{pq}\right)\right]\cr
&&\qquad\qquad{}=-\int d\tau\,d^D\bx\,\nabla_iN_j\frac{\delta W}{\delta g_{ij}}
=-\int d\tau\,d^D\bx\,\p_i\left(N_j\frac{\delta W}{\delta g_{ij}}\right),
\eea
as a consequence of the Bianchi identity $\nabla_i (R^{ij}-Rg^{ij}/2)=0$,   
or alternatively as a consequence of the gauge invariance of $W$ under 
spatial diffeomorphisms.  

Introducing an auxiliary field $B^{ij}$, we can rewrite (\ref{imact}) in the 
following form,
\bea
\label{baction}
S&=&\frac{i}{\kappa^2}\int d\tau\,d^D\bx\,\sqrt{g}N\left\{B^{ij}
\left[\frac{1}{N}\left(\dot g_{ij}-\nabla_iN_j-\nabla_jN_i
\right)+\frac{\kappa^2}{2\kappa_W^2}\CG_{ijk\ell}\left(R^{k\ell}-
\frac{1}{2}Rg^{k\ell}\right)\right]\right.\cr
&&\qquad\qquad\qquad\qquad\left.{}-\frac{1}{2}B^{ij}\CG_{ijk\ell}B^{k\ell}
\right\}.
\eea
All terms in (\ref{baction}) are at least linear in $B^{ij}$.  
This is a hallmark of similar constructions in nonequilibrium dynamics 
\cite{lebellac}, dynamical critical phenomena \cite{ma,halperin}, quantum 
critical systems \cite{sachdev,ardonne} and stochastic quantization 
\cite{parisi,namiki,zinnzwan}.  Moreover, the coefficient of the 
term linear in $B^{ij}$ has a special form, intimately related to an evolution 
equation for $g_{ij}$, 
\bea
\label{flw}
\dot g_{ij}&=&-\frac{\kappa^2}{2\kappa_W^2}N\CG_{ijk\ell}\left(R^{k\ell}-
\frac{1}{2}Rg^{k\ell}\right)+\nabla_iN_j+\nabla_jN_i
\dot g_{ij}\cr
&&\quad{}\equiv\frac{\kappa^2}{2}N\CG_{ijk\ell}
\frac{\delta W}{\delta g_{k\ell}}+\nabla_iN_j+\nabla_jN_i.
\eea
Since the curvature terms in this equation originated from the variational 
principle, this equation simply states that the evolution of $g_{ij}$ is 
governed by a gradient flow $\delta W/\delta g_{ij}$ on the space of metrics, 
up to possible gauge transformations represented by $N_i$ and $N$.  In the 
context of condensed matter applications mentioned above, systems whose action 
$S$ is so associated with a gradient flow generated by some $W$ are said to 
satisfy the condition of ``detailed balance.''  Investigating under what 
circumstances quantum corrections preserve these features of the action is the 
key to proving renormalizability of this setup.  

Under rather general circumstances, theories which satisfy the detailed 
balance condition have simpler quantum properties than a generic theory in 
$D+1$ dimensions.  Their renormalization properties are often inherited from 
the simpler renormalization of the associated theory in $D$ dimensions with 
action $W$, plus the possible renormalization of the relative normalization 
between the kinetic and potential terms in $S$.  Examples of this phenomenon 
include scalar fields \cite{zinn} or Yang-Mills gauge theories \cite{zinnzwan} 
(see also \cite{cym}).  It will be important to analyze under what 
circumstances an analog of such ``quantum inheritance principle'' is valid for 
our nonrelativistic gravity models.  This analysis is, however, beyond the 
scope of the present paper.

In passing, we note that the structure of the evolution equation (\ref{flw}) 
suggests an intimate relation between our theory of nonrelativistic gravity 
and the theory of Ricci flows, which in turn play a central role in Perelman's 
approach \cite{perelman} to the Poincar\'e conjecture.  Indeed, (\ref{flw}) 
is a covariantized Ricci flow equation, or more precisely a family of 
generalized Ricci flows parametrized by $\lambda$,
\be
\label{rflw}
\dot g_{ij}=-\frac{\kappa^2}{2\kappa_W^2}N\left[R_{ij}+
\frac{1-2\lambda}{2(D\lambda-1)}Rg_{ij}\right]+\nabla_iN_j+\nabla_jN_i.
\ee
Setting $N=1$ and $N_i=0$ recovers the naive Ricci flow equation.  The 
decorations of the naive flow in (\ref{rflw}) by $N$ and $N_i$ take into 
account the fact that geometrically, we only care about the flow up to a 
-- possibly time-dependent -- spatial diffeomorphism and a time 
reparametrization.  These gauge symmetries of the Ricci flow problem match 
naturally the foliation-preserving diffeomorphism invariance of our gravity 
theory.

\subsection{Hamiltonian Formulation}
\label{secadm}

It is instructive to rewrite our theory with foliation-preserving 
diffeomorphism invariance in the canonical formalism, generalizing the ADM 
formulation of general relativity.  The Hamiltonian formulation is 
particularly natural for the class of gravity theories proposed here, 
because the $D+1$ split of the spacetime variables is naturally compatible 
with the preferred role of time and the anisotropic scaling.  

The canonical momenta conjugate to $g_{ij}$ are
\be
\label{canmom}
\pi^{ij}=\frac{\delta S}{\delta\dot g_{ij}}=\frac{\sqrt{g}}{\kappa^2N}
G^{ijk\ell}\left(\dot g_{k\ell}-\nabla_kN_\ell-\nabla_\ell N_k\right)=
\frac{2\sqrt{g}}{\kappa^2}G^{ijk\ell}K_{k\ell},
\ee
where
\be
K_{ij}=\frac{1}{2N}\left(\dot g_{ij}-\nabla_iN_j-\nabla_j N_i\right)
\ee
is the extrinsic curvature tensor on the spatial leaves of the spacetime 
foliation.  The momenta conjugate to $N$ and $N_i$ are identically zero, 
and their vanishing represent primary first-class constraints.    
The Poisson bracket of the canonical variables is
\be
[g_{ij}(\bx),\pi^{k\ell}(\by)]=\frac{1}{2}\left(\delta_i^k
\delta_j^\ell+\delta_i^\ell\delta_j^k\right)\delta^D(\bx-\by).
\ee
In terms of the canonical variables, the Hamiltonian takes the form 
\be
\label{hadm}
H=\int d^D\bx\,\left(N\CH+N_i\CH^i\right),
\ee
with $\CH$ and $\CH^i$ given by
\bea
\CH&=&\frac{\kappa^2}{2\sqrt{g}}\pi^{ij}\CG_{ijk\ell}\pi^{k\ell}+
\frac{\kappa^2\sqrt{g}}{4\kappa_W^4}\left(R^{ij}-\frac{1}{2}Rg^{ij}\right)
\CG_{ijk\ell}\left(R^{k\ell}-\frac{1}{2}Rg^{k\ell}\right)\cr
&=&\frac{\kappa^2}{2\sqrt{g}}\left[\pi^{ij}\pi_{ij}
-\frac{\lambda}{D\lambda -1}(\pi_i^i)^2\right]
+\frac{\kappa^2\sqrt{g}}{4\kappa_W^4}\left(R^{ij}R_{ij}+a R^2\right) 
\eea
(where $a$ is again as in (\ref{alpha})), and
\be
\label{momcon}
\CH^i=-2\nabla_j\pi^{ij}.  
\ee
We would now like to calculate the algebra satisfied by the constraints in 
our nonrelativistic theory.  

For comparison, it will be useful to recall first the structure of the 
relativistic constraints.  In general relativity formulated in the canonical 
ADM formalism, the Hamiltonian is also given by (\ref{hadm}).  The momentum 
constraints $\CH_i$ take the same form as given in (\ref{momcon}), while 
$\CH$ is replaced by the relativistic Hamiltonian constraint
\be
\label{hperp}
\CH_\perp=\frac{16\pi G_{\!N}^{}}{2\sqrt{g}}\pi^{ij}\CG_{ijk\ell}\pi^{k\ell}-
\frac{\sqrt{g}}{16\pi G_{\!N}^{}}\left(R-2\Lambda\right),
\ee
where $G_{\!N}^{}$ is the Newton constant, and $\lambda$ has been set equal to 
$1$.  The quantum version of this constraint yields the Wheeler-De~Witt 
equation.  

General relativity is fundamentally built on the principle of spacetime 
diffeomorphism invariance.  One might therefore expect that the fist-class 
constraints $\CH_i(t,\bx)$ and $\CH_\perp(t,\bx)$ just confirm the naive 
expectation, and form the algebra of spacetime diffeomorphisms.  However, 
as is well-known, it is not so:  Under the Poisson bracket, the constraints 
of general relativity do not even close to form a Lie algebra.  Their  
commutation relations are 
\bea
\label{grhh}
[\int d^D\bx\,\zeta(\bx)\CH_\perp(\bx),\int d^D\by\,\eta(\by)\CH_\perp(\by)]
&=&\int d^D\bx\,(\zeta\p_i\eta-\eta\p_i\zeta)\,g^{ij}\CH_j(\bx),\\
\label{grhih}
[\int d^D\bx\,\zeta^i(\bx)\CH_i(\bx),\int d^D\by\,\eta(\by)\CH_\perp(\by)]
&=&\int d^D\bx\,\zeta^i\p_i\eta\,\CH_\perp(\bx)\\
\label{grhihi}
[\int d^D\bx\,\zeta^i(\bx)\CH_i(\bx),\int d^D\by\,\eta^j(\by)\CH_j(\by)]
&=&\int d^D\bx\,(\zeta^i\p_i\eta^k-\eta^i\p_i\zeta^k)\,\CH_k(\bx).
\eea
First, (\ref{grhihi}) is easy to interpret:  It shows that the $\CH_i$ 
constraints form the Lie algebra of generators of spatial diffeomorphisms, 
preserving the time foliation of the canonical formalism.  Similarly, 
(\ref{grhih}) simply 
indicates that $\CH_\perp(\by)$ transforms as a density under the spatial 
diffeomorphisms generated by $\CH_i$.  The subtlety occurs in the commutation 
relation of two $\CH_\perp$:  Because of the explicit presence of $g^{ij}$ in 
(\ref{grhh}), the structure ``constants'' are field-dependent, and strictly 
speaking, the constraints do not form a Lie algebra.  This fact contributes to 
the notorious conceptual as well as technical difficulties in the process of 
quantization of the relativistic theory (see, \eg , \cite{kuchar,ishamtime}).  

In our nonrelativistic gravity, the structure of constraints is slightly 
different than in general relativity.  If the lapse field $N$ is restricted 
to a function of time only, the constraint algebra is generated by the 
momentum constraints $\CH_i(t,\bx)$, which take the general relativistic form 
(\ref{momcon}), and the integral of 
$\CH$: 
\be
\qquad \CH_0\equiv\int d^D\bx\,\CH(t,\bx).
\ee
It is easy to show that these constraints form a closed algebra.  The 
commutator of two $\CH_i(\bx)$ generators coincides with (\ref{grhihi}).  Our 
$\CH(\bx)$ transforms as a density under generators $\CH_i$ of spatial 
diffeomorphisms and therefore satisfies (\ref{grhih}), implying that 
$\CH_i(\bx)$ commute with the zero mode $\CH_0$, as can be seen by setting 
$\eta=1$ in (\ref{grhih}).  (For this to work, it is important that 
$\CH(\bx)$ transforms as a density, which eliminates possible terms 
$\sim \eta\p_i\zeta^i$ in (\ref{grhih}).)  Finally, $\CH_0$ of course commutes 
with itself.  

The general theory of constrained systems \cite{ht} can be used to predict the 
number of physical degrees of freedom in our system.  There are $2D$ 
first-class constraints per spacetime point: $D$ components of 
$\CH_i$ and $D$ momenta conjugate to $N_i$.  We also have $D(D+3)$ fields: 
$D(D+1)/2$ components of $g_{ij}$ and their conjugate momenta, and $D$ 
components of $N_i$ and their momenta.  The expected number of degrees 
of freedom per spacetime point is
\bea
\#({\rm DoF})&=&\frac{1}{2}\left(\vphantom{X^i}\#({\rm field\ components})
-2\times\#({\rm first\mbox{-}class\ constraints})\right)={}\cr
&&\qquad\qquad{}=\frac{D(D-1)}{2}=\frac{(D+1)(D-2)}{2}+1.
\eea
The number of massless graviton polarizations in relativistic gravity in 
$D+1$ spacetime dimensions is $(D+1)(D-2)/2$.  Thus, compared to general 
relativity, our theory is generically expected to have one additional 
propagating scalar degree of freedom, at least in the absence of any 
additional gauge symmetry. 

\subsection{At the Free-Field Fixed Point with $z=2$}
\label{secffpt}

In order to prepare for the study of the full interacting theory, it is 
useful to first understand the properties of its free-field fixed point 
limit.  Free-field limits of anisotropic theories with nontrivial dynamical 
critical exponent $z$ exhibit interesting properties, such as families of 
inequivalent fixed points, as we have seen in the example of $z=2$ Yang-Mills 
theory in \cite{cym}.

\subsubsection{Scaling Properties and the Critical Dimension}

By design, our nonrelativistic gravity has a free-field limit with anisotropic 
scaling of space and time, characterized by dynamical critical exponent 
$z=2$.  The engineering dimensions (\ie , the scaling dimensions at the $z=2$ 
free-field fixed point) of various quantities are as follows.  First, just as 
in general relativity, the metric components $g_{ij}$ are naturally 
dimensionless as a result of their geometric origin. The dimensions of the 
remaining fields are then determined to be
\be
[g_{ij}]=0,\qquad [N_i]=1,\qquad [N]=0.
\ee
In the formulation that uses the auxiliary field $B^{ij}$, we also have 
$[B^{ij}]=2$.  

The coupling constants appearing in (\ref{fullact}) have dimensions
\be
[\kappa]=\frac{2-D}{2},\qquad [\kappa_W^{}]=\frac{2-D}{2},\qquad [\lambda]=0.
\ee
As in the system of the Lifshitz scalar at $z=2$, making the gravity theory 
anisotropic with dynamical exponent $z=2$ has shifted the critical dimension 
of the free-field fixed point,  from $1+1$ to $2+1$.  This is the dimension 
where both $\kappa$ and $\kappa_W^{}$ are dimensionless.  Of course, in the 
critical dimension $D=2$, the Einstein tensor and consequently the potential 
term $S_K$ in the action vanish identically.  This simplification of $z=2$ 
gravity in the critical case of $2+1$ dimensions is closely related to the 
simplification of relativistic gravity in $1+1$ dimensions, where the 
Einstein-Hilbert action is a topological invariant.  

The free-field fixed point is defined by ``turning off'' all the coupling 
constants that measure interactions.  Our theory has three couplings:  
$\kappa_W^{}$, $\kappa$ and $\lambda$.  As it turns out, only one of them  
measures the strength of self-interactions of the gravitons, and turning it 
off makes the theory free.  More precisely, turning off the interactions is 
equivalent to sending $\kappa_W^{}$ to zero while keeping $\lambda$ and the 
ratio 
\be
\gamma=\frac{\kappa}{\kappa_W^{}}
\ee
fixed.  This leaves two dimensionless coupling constants $\gamma$ and 
$\lambda$ which survive in the noninteracting limit and measure the 
properties of the free-field fixed point.  Thus, we obtain a two-parameter 
family of fixed points, all with $z=2$.  This is very analogous to the case of 
quantum critical Yang-Mills theory studied in \cite{cym}, which exhibits 
a similar one-parameter family of free fixed points with $z=2$.

\subsubsection{The Spectrum}
\label{secspec}

We will now determine the spectrum of physical excitations, and their 
dispersion relations, in the family of free fixed point parametrized by 
$\gamma$ and $\lambda$.  

The action at the free-field fixed point can be found by expanding the 
theory around the flat background with $g_{ij}=\delta_{ij}$, $N=1$ and 
$N_i=0$.  This background is indeed a classical solution of the theory, 
for any value of $\gamma$ and $\lambda$.  We expand the fields around 
this solution, writing
\be
g_{ij}=\delta_{ij}+\kappa_W^{}h_{ij}.
\ee
$N_i$ are of order $\kappa_W^{}$, and we rescale them accordingly.  Finally, 
the corrections to $N=1$ drop out in this approximation.   

The Gaussian action of the linearized theory is then
\bea
\label{quadact}
S&=&\frac{1}{2}\int dt\,d^D\bx\left\{\frac{1}{\gamma^2}\left[
\left(\dot h_{ij}-\p_iN_j-\p_jN_i\right)
\left(\dot h_{ij}-\p_iN_j-\p_jN_i\right)
-\lambda\left(\dot h_{ii}-2\p_iN_i\right)^2\right]\right.\cr
&&{}-\frac{\gamma^2}{16}\,h_{ij}\left[
\frac{(D-2)(2\lambda-1)}{D\lambda-1}\left(\p_i\p_j\p_k\p_\ell+\delta_{ij}
\delta_{k\ell}\left(\p^2\right)^2-2\delta_{ij}\p_k\p_\ell\p^2\right)\right.\cr
&&\qquad\left.\left.{}+2\left(\delta_{ij}\p_k\p_\ell-\delta_{ik}\p_j\p_\ell
\right)\p^2+(\delta_{ik}\delta_{j\ell}-\delta_{ij}\delta_{k\ell})\left(\p^2
\right)^2\vphantom{\frac{1}{2}}\right]h_{k\ell}\right\}.
\eea
In order to identify the propagating modes and determine their dispersion 
relations, we must make a suitable gauge choice and diagonalize this action.  
Given the nonrelativistic character of the theory, it is natural to 
choose 
\be
\label{ngauge}
N_i=0
\ee
as our gauge-fixing condition.  This gauge choice does not fix the gauge 
symmetries completely, leaving the group of time-independent spatial 
diffeomorphisms unfixed.  In addition, making this gauge choice implies that 
the fields in (\ref{quadact}) are constrained by the following analog of the 
Gauss constraint, 
\be
\label{lgauss}
\p_i \dot h_{ij}=\lambda\p_j\dot h,
\ee
where $h\equiv h_{ii}$.  This constraint comes from the linearized equation of 
motion of $N_i$ in the full gauge-invariant action.  We can fix the residual 
gauge symmetry by setting 
\be
\label{resigf}
\p_i h_{ij}-\lambda\p_j h=0,
\ee
at some fixed time surface, $t=t_0$.  The constraint (\ref{lgauss}) then 
implies that (\ref{resigf}) continues to hold at all times.  

(\ref{resigf}) is a legitimate gauge choice for values of $\lambda$ not equal 
to one.  When $\lambda=1$, (\ref{resigf}) is not attainable by a spatial 
diffeomorphism.  The simplest way to see that is to apply $\p_j$ to 
(\ref{resigf}).  The left hand side then equals $\p_j\p_ih_{ij}-\p^2h=R$, the 
linearized Ricci scalar which cannot be set to zero by a gauge 
transformation.  As we will see below, $\lambda=1$ is indeed a special 
case, where the free-field fixed point exhibits an enhanced gauge symmetry.  

In order to read off the physical polarizations of the metric and their 
dispersion relations, we need to rewrite the quadratic action (\ref{quadact}) 
in variables that automatically take into account the constraints 
(\ref{lgauss}).  We switch from $h_{ij}$ to the new variables, defined as 
\be
H_{ij}=h_{ij}-\lambda\delta_{ij}h.  
\ee
Our residual gauge fixing (\ref{resigf}) together with the Gauss constraint 
(\ref{lgauss}) imply that $H_{ij}$ is transverse, 
\be
\p_iH_{ij}=0.
\ee
The transverse tensor $H_{ij}$ contains all the physical polarizations of the 
metric.  In order to separate the individual modes, we further decompose 
$H_{ij}$ into the transverse traceless part $\tilde H_{ij}$ and the trace part 
$H$:
\be
H_{ij} =\tilde H_{ij}+\frac{1}{D-1}\left(\delta_{ij}-\frac{\p_i\p_j}{\p^2}
\right)H.
\ee
We have $\tilde H_{ii}=0$, $\p_i\tilde H_{ij}=0$, and $H=H_{ii}$.    

In this gauge, the linear equations of motion that follow from (\ref{quadact}) 
can be diagonalized, and one can determine the number of physical 
polarizations and their dispersion relations.  $\tilde H_{ij}$ yields 
$(D-2)(D+1)/2$ transverse traceless polarizations, all with the same 
dispersion relation
\be
\label{drhi}
\omega^2=\frac{\gamma^4}{16}(\bk^2)^2.  
\ee
In addition, the trace $H$ leads to one mode, whose dispersion relation is
\be
\label{drh}
\omega^2=\frac{\Gamma^4}{16}(\bk^2)^2,
\ee
with
\be
\Gamma^4=\frac{(D-2)^2(\lambda-1)^2}{(D\lambda-1)^2}\gamma^4.
\ee

The free-field fixed point is well-defined in the large range of the 
parameters $\lambda$ and $D$ for which the energy of the excitations is 
bounded from below.  We can identify this physical range of parameters  
by expressing the gauge-fixed action in terms of $\tilde H_{ij}$ and $H$.  
The kinetic term becomes
\be
S_K=\frac{1}{2\gamma^2}\int dt\,d^D\bx\left\{\dot{\tilde H}_{ij}
\dot{\tilde H}_{ij}+\frac{\lambda-1}{(D-1)(D\lambda-1)}\dot H^2\right\},
\ee
and the potential term is
\be
S_V=\frac{1}{32\gamma^2}\int dt\,d^D\bx\left\{\p^2\tilde H_{ij}\p^2
\tilde H_{ij}+\frac{(D-2)^2(\lambda-1)^3}{(D-1)(D\lambda-1)^3}\left(\p^2H
\right)^2\right\}.
\ee
Hence, assuming $D>1$ the energy of the physical modes is positive definite 
when $\lambda<1/D$ or $\lambda>1$.  In the complementary regime 
$1/D<\lambda<1$, the scalar mode $H$ is a ghost.  

The dispersion relation (\ref{drh}) for the scalar mode $H$ suggests that 
something special happens at $\lambda=1/D$ and $\lambda=1$.  When 
$\lambda=1/D$, the De~Witt metric develops a null direction.  As a result, 
this is the value at which the theory may develop a local version of conformal 
symmetry, depending on the specific form of the potential term in the action. 
This case will be relevant to our membrane theory in Section~\ref{secmemc}, 
where we will be interested in incorporating a local Weyl invariance in 
$2+1$ dimensions.  

At the other special value, $\lambda=1$, the equation of motion for 
the scalar mode $H(t,\bx)$ collapses to $\ddot{H}=0$, with the general 
solution
\be
\label{hgen}
H(t,\bx)=H_0(\bx)+tH_1(\bx).
\ee
If present, such degrees of freedom would be difficult to interpret as 
physical excitations.  As it turns out, at this value of $\lambda$, the 
linearized theory develops an enhanced gauge symmetry, acting via
\be
\label{reduone}
\delta N_i=\p_i\varepsilon(\bx),\qquad \delta h_{ij}=0,
\ee
\ie , as a time-independent $U(1)$ gauge transformation.  This is 
the Abelian symmetry (\ref{uone}), linearized and reduced to preserve $A=0$.  
This spatial gauge symmetry plays an interesting role in the theory.  The 
$N_i=0$ gauge can now be attained in two steps, first by using a 
diffeomorphism to get $N_i=\p_i u$ for some function $u(\bx)$, and then using 
(\ref{reduone}) with $\varepsilon=-u$ to set $N_i=0$.  The first step leaves 
an extra unfixed diffeomorphism symmetry, given by $\zeta^i(t,\bx)$ that 
satisfy $\dot\zeta^i=\p_iu(\bx)$.   The generators of such unfixed 
diffeomorphisms are of the form 
\be
\zeta^i(t,\bx)=\zeta^i_0(\bx)+t\p_iu(\bx).
\ee
These residual diffeomorphisms acts on $H$ via $\delta H\sim \p_i\zeta^i$.  
The extra gauge freedom given by $u(\bx)$ can be used to set $H_1$ in 
(\ref{hgen}) to zero, leaving the transverse traceless gravitons as the only 
physical excitations at $\lambda=1$.   

\subsection{Relevant Deformations: Lower-Dimension Operators in the Potential 
Term}
\label{seclowd}

So far, we concentrated on the terms in the action which have the same 
engineering dimension as the kinetic term (\ref{kin}).  These are the terms 
that determine the behavior of the $z=2$ fixed point.  Now we extend our 
analysis to incorporate operators with lower dimensions, compatible with the 
symmetries of the theory.  If such operators exist, general arguments from 
effective field theory indicate that such terms will be generated by quantum 
effects, and will dominate over the original terms in $S_V$ in the 
long-distance dynamics of the theory.  

We will discuss the issue of relevant deformations of $z=2$ gravity in 
$D+1$ dimensions only briefly, because in Section~\ref{secmemc} we will follow 
a different route:  We will impose an additional gauge symmetry, related to 
Weyl invariance, which will forbid any lower-dimensional operators in $z=2$ 
gravity in $2+1$ dimensions.

In theories satisfying the detailed balance condition, there is a hierarchy of 
ways in which lower-dimension operators can be added to the classical theory:
\begin{enumerate}
\item  In the minimal modification, we add lower-dimensional operators to 
$W$, and thus preserve the detailed balance condition. 
\item We can add terms to $E^{ij}$ which respect all the symmetries but 
cannot be derived from varying any action in $D$ dimensions (if such terms 
exist).%
\footnote{In fact, this can be done already for terms of the same dimension as 
those in $E^{ij}$.  For example, in the theory of a single Lifshitz 
scalar reviewed in Section~\ref{seclifs}, an example of such a term is 
$\p_i\Phi\p_i\Phi$.  The addition of this term changes the theory 
 radically, from the Lifshitz theory to the universality class associated 
with the KPZ equation known from the nonequilibrium problem of surface growth.}
\item Finally, one can simply add lower-dimension operators directly to the 
action in $D+1$ dimensions, softly breaking not only the condition of detailed 
balance, but also the fact that in the representation with the auxiliary field 
$B$, only terms at least linear in $B$ appear in the action.  
\end{enumerate}

In the following, we will mostly focus on the first option, in which 
lower-dimensional terms are added to $W$.  For $z=2$ gravity without matter, 
the only such term that can be added to $W$ is the cosmological constant 
term.  Restoring the cosmological constant in (\ref{ehact}), 
\be
W=\frac{1}{\kappa_W^2}\int d^D\bx\,\sqrt{g}(R-2\LambdaD),
\ee
we get the following action in $D+1$ dimensions 
\bea
S&=&\frac{1}{2}\int dt\,d^D\bx\,\sqrt{g}\left\{\frac{1}{\kappa^2N}
\left(\dot g_{ij}-\nabla_iN_j-\nabla_jN_i\right)G^{ijk\ell}
\left(\dot g_{k\ell}-\nabla_kN_\ell-\nabla_\ell N_k\right)
\right.\nonumber\\
&&\left.\qquad{}-\frac{\kappa^2N}{4\kappa_W^4}\left(R^{ij}-\frac{1}{2}Rg^{ij}
+\LambdaD g^{ij}\right)\CG_{ijk\ell}
\left(R^{k\ell}-\frac{1}{2}Rg^{k\ell}+\LambdaD g^{k\ell}\right)
\right\}.
\eea
In dimensions $D>2$, turning on $\LambdaD$ in $W$ induces two new terms 
in $S_V$: The spatial Ricci scalar term $R$ and the spatial volume term.  
In $2+1$ dimensions, since $R^{ij}-Rg^{ij}/2$ vanishes identically, no Ricci 
scalar term is produced in $S_V$.  We will return to this case in detail 
in Section~\ref{secmemc}, and limit our present discussion to $D>2$.  

It is natural to define a scale $M$, 
\be
M^2=\frac{D-2}{1-D\lambda}\LambdaD.  
\ee
In terms of $M$, the action becomes
\bea
\label{maction}
S&=&\frac{1}{2}\int dt\,d^D\bx\,\sqrt{g}\left\{\frac{1}{\kappa^2N}
\left(\dot g_{ij}-\nabla_iN_j-\nabla_jN_i\right)G^{ijk\ell}
\left(\dot g_{k\ell}-\nabla_kN_\ell-\nabla_\ell N_k\right)
\right.\nonumber\\
&&\left.\qquad\qquad{}-\frac{\kappa^2N}{4\kappa_W^4}\left(R^{ij}R_{ij}
+a R^2-M^2 R+\frac{D(1-D\lambda)}{(D-2)^2}M^4\right)\right\}.
\eea
The constant $a$ takes the value given in (\ref{alpha}).  

For simplicity, we will assume $M^2>0$.  If $\lambda>1/D$, this 
means starting with a negative cosmological constant, $\LambdaD<0$, in $W$.  
The other sign of $M^2$ would correspond to the gravity analog 
of the ``spatially modulated'' phases in the Lifshitz scalar theory, which we 
will not study in this paper.  

Under the influence of the deformation by lower-dimension operators, the 
theory will flow from $z=2$ at short distances, to $z=1$ in the infrared.  
This flow to $z=1$ is in fact generic for quantum theories of the Lifshitz 
type (see \cite{cym}).  The dynamics of the theory at long distances will 
be dominated by the most relevant operators.  In our gravity theory, those 
will be the terms in $S_V$ with couplings involving nonzero powers of $M$:  
The spatial Ricci scalar, and the spatial volume term.   Together with the 
kinetic term $S_K$, these are exactly the ingredients that are required in 
general relativity.  

In order to compare the long-distance physics of the theory deformed by 
relevant operators to that of Einstein's theory, it is natural to redefine 
the time coordinate,
\be
\label{ceff}
x^0=ct,\qquad c=\frac{\gamma^2}{4}M.
\ee
This is one of the most notable features of our construction: The effective 
long-distance speed of light originates microscopically from a relevant 
coupling in the theory describing the anisotropic short-distance dynamics.  

In these relativistic coordinates, the dominant long-distance terms in the 
Hamiltonian $\CH(\bx)$ of our deformed theory are precisely such that they 
reproduce the relativistic Hamiltonian $\CH_\perp(\bx)$ of (\ref{hperp}), 
with the effective Newton constant 
\be
\label{geff}
G_N=\frac{\kappa_W^2}{8\pi M},
\ee
and the effective cosmological constant 
\be
\label{leff}
2\Lambda=\frac{D(1-D\lambda)}{(D-2)^2}M^2=\frac{D}{D-2}\LambdaD.
\ee
Thus, we conclude that
\begin{itemize}
\item under the influence of relevant deformations, the anisotropic gravity 
theory flows in the infrared limit naturally to a theory with isotropic 
scaling and $z=1$, and leads to long-distance physics which is remarkably 
close to general relativity.  
\item There are several differences between the general relativity and the 
$z=1$ infrared limit of our theory.  First, our Hamiltonian depends on the 
additional coupling $\lambda$, which equals 1 in general relativity.  
In addition, we restricted the lapse variable $N$ to be independent of 
spatial coordinates.  
\item Notably, the emerging long-distance speed of light (\ref{ceff}), the 
effective Newton constant (\ref{geff}), and also the effective cosmological 
constant (\ref{leff}) all originate from the relevant deformations of a 
deeply nonrelativistic short-distance theory of gravity with anisotropic 
scaling and $z=2$.  
\item While interactions and quantum effects will affect some features of the 
flow, our conclusions are exact in the noninteracting limit $\kappa_W^{}=0$.  
\end{itemize}

\section{Membranes at Criticality: $z=2$ Gravity and Matter in $2+1$ 
Dimensions}
\label{secmemc}

Having presented the construction of $z=2$ gravity, we can now return to 
our original problem, and combine this theory in $2+1$ dimensions with 
Lifshitz matter, in order to establish the desired connection to the 
partition function of the bosonic string.  In the process, we must clarify 
how the worldsheet Weyl invariance of critical string theory can be 
incorporated into our scheme.  

\subsection{Coupling $z=2$ Gravity to Lifshitz Matter} 

We now consider the $z=2$ gravity theory in its critical dimension $2+1$, 
coupled to 26 Lifshitz scalar fields $X^I(t,\bx)$, $I=1,\ldots 26$.  
Our starting point is the Polyakov worldsheet action for the bosonic string of 
Euclidean worldsheet signature, embedded in the spacetime target manifold 
$\R^{26}$ parametrized by coordinates $X^I$ and equipped with the flat 
Euclidean metric $\delta_{IJ}$:  
\be
\label{polyakov}
W=\frac{1}{4\pi\alpha'}\int d^2\bx\,\sqrt{g}\,g^{ij}\p_iX^I\p_jX^I.
\ee
Combining the construction of $z=2$ gravity presented in 
Section~\ref{seclifgrav} with the Lifshitz matter reviewed in 
Section~\ref{seclifs}, we obtain the action of the coupled system of $z=2$ 
gravity and $z=2$ matter in $2+1$ dimensions, 

\bea
\label{crmembact}
S&=&\frac{1}{2}\int dt\,d^2\bx\,\sqrt{g}\left\{\frac{1}{\kappa^2N}
\left(\dot g_{ij}-\nabla_iN_j-\nabla_jN_i\right)G^{ijk\ell}
\left(\dot g_{k\ell}-\nabla_kN_\ell-\nabla_\ell N_k\right)
\right.\nonumber\\
&&\left.\qquad{}+\frac{1}{\alpham N}\left(\dot X-N^i\p_iX\right)^2
-\frac{\alpham\,N}{(4\pi\alpha')^2}(\Delta X)^2
-\frac{\kappa^2N}{4(4\pi\alpha')^2}T^{ij}\,\CG_{ijk\ell}\,T^{k\ell}\right\}.  
\eea
Here 
\be
\Delta X^I\equiv\frac{1}{\sqrt{g}}\p_i(\sqrt{g}g^{ij}\p_jX^I)
\ee
is the Laplace operator of $g_{ij}$ acting on the scalar field $X^I$, and 
\be
T_{ij}\equiv 4\pi\alpha'\frac{1}{\sqrt{g}}\frac{\delta W}{\delta g^{ij}}=
\p_iX^I\p_jX^I-\frac{1}{2}g_{ij}(g^{k\ell}\p_kX^I\p_\ell X^I)
\ee
is the energy-momentum tensor of the scalar fields in (\ref{polyakov}).  
This action is gauge invariant under foliation-preserving diffeomorphisms, 
and satisfies the detailed balance condition with respect to 
(\ref{polyakov}).   Under the foliation-preserving diffeomorphisms 
(\ref{fdif}), the Lifshitz scalars transform as
\be
\delta X^I=f\,\dot X^I+\zeta^i\p_iX^I.
\ee

Note that the kinetic term for the Lifshitz scalars in (\ref{crmembact}) 
required the introduction of a new coupling $\alpham$ of the same dimension as 
$\alpha'$, \ie , spacetime length squared.  We can express this new spacetime 
length scale in terms of $\alpha'$, and define a new dimensionless parameter 
\be
\kappa_M^{2}=\frac{\alpham}{4\pi\alpha'}
\ee
instead.  In addition to $\kappa_M^{}$, the theory has two other dimensionless 
couplings: $\lambda$, which is hidden in the definition of the De~Witt metric 
(\ref{dwtt}), and $\kappa$.  Since the coupling of matter to gravity leads to 
the nonlinear terms $T_{ij}T^{ij}$ in (\ref{crmembact}), the theory is 
interacting at nonzero $\kappa$.  The free limit corresponds to taking 
$\alpha'\to 0$ and $\kappa\to 0$.  The remaining two dimensionless 
parameters $\kappa_M^{}$ and $\lambda$ survive in the free field limit, and 
characterize the properties of the family of free-field fixed points in $z=2$ 
gravity with matter in $2+1$ dimensions.  

\subsection{Anisotropic Weyl Symmetry}

The gauge symmetries of our coupled theory (\ref{crmembact}) do not yet match 
those of critical string theory.  Upon canonical quantization, worldsheet 
diffeomorphisms are reproduced as symmetries of wavefunctions, but Weyl 
invariance is not.  

One could go in the direction of noncritical string theory, and try to 
develop a corresponding noncritical theory of membranes.%
\footnote{This could be relevant for the relation between noncritical strings 
in two dimensions, noncritical M-theory in $2+1$ dimensions, and topological 
strings of the A-model, as discussed in \cite{pixi2}.   This possibility 
was indeed one of the original motivations for this project.}
In this paper, we are more interested in reproducing the conventional 
critical bosonic string, and we must therefore look for an implementation 
of a $2+1$ dimensional analog of Weyl invariance on the membrane 
worldvolume, as an additional gauge symmetry supplementing the foliated 
diffeomorphisms.  

The requirement of a local Weyl invariance will actually fix the value of 
$\lambda$ of the gravity sector uniquely.  Moreover, this gauge invariance 
extends to the matter sector as well, described by the Lifshitz scalar 
theory.  We define the ``anisotropic Weyl transformations'' -- for any value 
of the dynamical critical exponent $z$ -- as follows,
\be
g_{ij}\rightarrow\exp\left\{2\Omega(t,{\bf x})\right\}g_{ij},\qquad
N_i\rightarrow\exp\left\{2\Omega(t,{\bf x})\right\}N_i,\qquad
N\rightarrow\exp\left\{z\Omega(t,{\bf x})\right\}N.
\ee
Since the anisotropic Weyl transformations act nontrivially on $N$, we can 
no longer restrict $N$ to be independent of space; $N$ is now a $2+1$ 
dimensional field, $N(t,\bx)$.  

Such anisotropic Weyl transformations with fixed $z$ form a closed algebra 
with foliation-preserving diffeomorphisms (\ref{foldif}):  Denoting by 
$\delta_\omega$ the infinitesimal Weyl transformation with parameter 
$\omega(t,\bx)$, and by $\delta_v$ the infinitesimal foliation-preserving 
diffeomorphism transformation $v\equiv(f(t),\zeta^i(t,\bx))$ as given in 
(\ref{foldif}), one can show that their commutator yields another anisotropic 
Weyl transformation, 
\be
[\delta_v,\delta_\omega]=\delta_{\tilde\omega},\qquad{\rm with}\ 
\tilde\omega=\zeta^i\p_i\omega+f\,\dot\omega.
\ee

Specializing to $z=2$, our Lagrangian is classically invariant under the 
anisotropic Weyl transformations if we set $\lambda=1/2$.  In the proof of 
this gauge invariance, it is important that for infinitesimal Weyl 
transformation $\omega$ (and again temporarily restoring arbitrary $z$ and 
arbitrary space dimension $D$ for future reference)
\be
\delta_\omega(\nabla_iN_j)=2\omega\nabla_iN_j
+N_j\p_i\omega-N_i\p_j\omega+g_{ij}g^{k\ell}N_k\p_\ell\omega.
\ee
Contracting this with the De~Witt metric, we get
\be
G^{ijk\ell}\delta_\omega(\nabla_iN_j)=2\omega\,G^{ijk\ell}\nabla_iN_j
+(1-D\lambda)g^{k\ell}g^{ij}N_i\p_j\omega.
\ee
Hence, for the conformal value $\lambda=1/D$, the terms with derivatives 
of $\delta\omega$ vanish, and the kinetic term for gravity will be invariant 
under the local anisotropic Weyl transformations.  

Returning now to the case of interest, $D=2$, we set the coupling constant 
$\lambda$ in the De~Witt metric $G^{ijk\ell}$ equal to its conformal value 
$\lambda=1/2$.  Our action (\ref{crmembact}) is gauge invariant under local 
anisotropic Weyl transformations, at least at the classical level.  This 
continues to be the case after coupling to the Lifshitz scalars $X^I$, 
provided they transform with weight zero under the Weyl transformations, 
$\delta_\omega X^I=0$.  The requirement of local anisotropic Weyl symmetry 
forbids any relevant terms in the action of our coupled system of $z=2$ 
gravity and $z=2$ matter in $2+1$ dimensions.  

\subsection{Canonical Formulation}

In order to understand properties of the ground-state wavefunction, we would 
like to quantize our $2+1$ dimensonal theory with anisotropic Weyl invariance 
canonically on $\Sigma_h\times\R$, where $\Sigma_h$ is the Riemann surface of 
genus $h$.  

We use the ADM formulation of Section~\ref{secadm}, generalized to the 
presence of matter.  The momenta conjugate to $g_{ij}$ were found in 
(\ref{canmom}):  
\be
\label{conjmm}
\pi^{ij}=\frac{\sqrt{g}}{\kappa^2N}
G^{ijk\ell}\left(\dot g_{k\ell}-\nabla_k N_\ell-\nabla_\ell N_k\right)
=\frac{2\sqrt{g}}{\kappa^2}G^{ijk\ell}K_{k\ell}.\ee
Once we set $\lambda=1/2$ in the De~Witt metric, as required by 
anisotropic Weyl invariance, we find that the momenta (\ref{conjmm}) 
\be
\pi^{ij}=\frac{2\sqrt{g}}{\kappa^2}\left(K^{ij}-\frac{1}{2}g^{ij}K\right),
\ee
(where $K\equiv g^{ij}K_{ij}$) are traceless,
\be
\label{tracelesspi}
\pi_i^i\equiv g_{ij}\pi^{ij}=0,
\ee
as a consequence of the local Weyl symmetry.  (\ref{tracelesspi}) is a new 
primary constraint, When this constraint is solved, only the traceless 
momenta -- which we denote by $\tilde\pi^{ij}$ -- appear in the theory.    

Similarly, the momenta $P_I$ conjugate to the Lifshitz scalars $X^I$ are
\be
P_I=\frac{\delta S}{\delta\dot X^I}=\frac{\sqrt{g}}{\alpham\,N}(\dot X^I
-N^i\p_iX^I).
\ee
The Hamiltonian is 
\be
\label{wham}
H=\int d^2\bx \left(N\CH+N_i\CH^i\right),
\ee
with 
\be
\CH=\frac{\kappa^2}{2\sqrt{g}}\tilde\pi^{ij}g_{ik}g_{j\ell}\tilde\pi^{k\ell}+
\frac{\alpham}{2\sqrt{g}}P_IP_I+\frac{\sqrt{g}}{2(4\pi\alpha')^2}
\left(\alpham\Delta X^I\Delta X^I+\frac{\kappa^2}{4}T_{ij}T^{ij}\right),
\ee
and
\be
\CH^i=-2\nabla_j\pi^{ij}+g^{ij}P_I\,\p_jX^I.
\ee

\subsection{The Algebra of Constraints}

The anisotropic Weyl invariance requires $N$ to be a general function of 
$t$ and $\bx$.  As a result, the structure of the Hamiltonian (\ref{wham}) 
indicates that in the Weyl invariant theory, both $\CH_i(\bx)$ and 
$\CH(\bx)$ (and not just its zero mode $\CH_0=\int d^2\bx\,\CH(\bx)$) will 
play the role of the constraints, and we must determine their algebra.    

As an alternative to the general relativistic constraints 
(\ref{grhh}-\ref{grhihi}), another algebra of ``general covariance'' was 
proposed in \cite{teitelboim}:
\bea
\label{ughh}
[\int d^D\bx\,\zeta(\bx)\CH_\perp(\bx),\int d^D\by\,\eta(\by)\CH_\perp(\by)]
&=&0,\\
\label{ughih}
[\int d^D\bx\,\zeta^i(\bx)\CH_i(\bx),\int d^D\by\,\eta(\by)\CH_\perp(\by)]
&=&\int d^D\bx\,\zeta^i\p_i\eta\,\CH_\perp(\bx),\\
\label{ughihi}
[\int d^D\bx\,\zeta^i(\bx)\CH_i(\bx),\int d^D\by\,\eta^j(\by)\CH_j(\by)]
&=&\int d^D\bx\,(\zeta^i\p_i\eta^k-\eta^i\p_i\zeta^k)\,\CH_k(\bx).
\eea
This in some sense is a nicer symmetry than (\ref{grhh}-\ref{grhihi}):  It 
actually forms a Lie algebra, with structure constants independent of the 
fields.  It is a symmetry of the so-called ``ultralocal theory of gravity'' 
\cite{teitelboim,isham} which in fact fits naturally into our framework: The 
action in the ultralocal theory of gravity also takes the form $S=S_K-S_V$, 
with $S_K$ given by (\ref{kin}), and $S_V$ set equal to zero 
(or to $\sqrt{g}\Lambda$ in the case of a nonzero cosmological constant).  

As it turns out, the ultralocal algebra (\ref{ughh}-\ref{ughihi}) is also 
the algebra of Hamiltonian constraints of $z=2$ gravity in $2+1$ dimensions 
with Weyl invariance and without matter.  The simplest way to see that is 
to notice that in the critical dimension $D=2$, the potential term $S_V$ 
in our $z=2$ theory vanishes identically, and the full action coincides 
that that of the ultralocal theory, with $\lambda=1/2$.

When Lifshitz matter is introduced, the commutator of two $\CH(\bx)$ no longer 
vanishes.  Instead, we get
\be
[\int d^2\bx\,\zeta(\bx)\CH(\bx),\int d^2\by\,\eta(\by)\CH(\by)]
=\int d^2\bx\,\left(\zeta\p_i\eta-\eta\p_i\zeta\right)\Phi^i(\bx),
\ee
with
\be
\Phi^i(\bx)=-\frac{\alpham}{(4\pi\alpha')^2}\left(\kappa^2\tilde\pi^{ij}\p_jX^I
\Delta X^I -\frac{\kappa^2}{2}P^I\p_jX^IT^{ij}+\alpham\, g^{ij}(P^I\p_j\Delta 
X^I-\p_jP^I\Delta X^I)\right).
\ee
One could attempt to add $\Phi^i(\bx)$ to the list of constraints, and 
continue the process until the constraint algebra closes.  However, there is a 
simpler alternative, which will go a long way towards solving our original 
problem.  Note first that $\Phi^i$ can be rewritten as
\bea
\Phi^i(\bx)&=&\frac{i\alpham}{(4\pi\alpha')^2}\left\{\kappa^2
\Delta X^I\p_jX^I\left(i\tilde\pi^{ij}-\frac{\sqrt{g}}{8\pi\alpha'}T^{ij}
\right)\right.\cr
&+&\left.\left(\alpham\, g^{ij}(\p_j\Delta X^I-\Delta X^I\p_j)
-\frac{\kappa^2}{2}T^{ij}\p_j 
X^I\right)\left(iP^I-\frac{\sqrt{g}}{4\pi\alpha'}\Delta X^I\right)\right\}.
\eea
This suggests introducing 
\bea
a^{ij}&=&i\pi^{ij}+\frac{1}{2}\frac{\delta W}{\delta g_{ij}}=
i\pi^{ij}-\frac{\sqrt{g}}{8\pi\alpha'}T^{ij},\\
Q^I&=&iP^I+\frac{1}{2}\frac{\delta W}{\delta X^I}=
iP^I-\frac{\sqrt{g}}{4\pi\alpha'}\Delta X^I,
\eea
and their complex conjugates 
\be
\bar a^{ij}=-i\pi^{ij}-\frac{\sqrt{g}}{8\pi\alpha'}T^{ij},\qquad
\bar Q^I=-iP^I-\frac{\sqrt{g}}{4\pi\alpha'}\Delta X^I.
\ee
In our system of gravity coupled to matter, $Q^I$ and $a^{ij}$ are the precise 
analogs of the $\CQ$ variable (\ref{cqvar}) defined in our discussion of the 
Lifshitz scalar theory in Section~\ref{seclifs}.  
In terms of these variables, the Hamiltonian constraint itself can be written 
as
\be
\CH=\frac{\kappa^2}{2\sqrt{g}}\bar a^{ij}G_{ijk\ell}\,a^{k\ell}+
\frac{\alpham}{2\sqrt{g}}\bar Q^IQ^I.
\ee
Given these facts, the following way towards quantization of the system 
suggests itself.  Instead of $\{\CH_i,\CH,\Phi^i,\ldots\}$, we can choose 
the constraints to be $\{\CH_i,a^{ij},Q^I\}$.  This may not be the unique 
possibility how to approach the quantization of our system, but it does 
exhibit the following attractive features:  
\begin{itemize}
\item
Since $\CH(\bx)$, $\CH_i(\bx)$ and $\Phi^i$ are linear in $a^{ij}$ and $Q^I$, 
the vanishing of our constraints $a^{ij}$ and $Q^I$ implies the vanishing 
of the Hamiltonian and momentum constraints $\CH$ and $\CH_i$, as well as 
$\Phi^i$.  Similarly, it implies that the constraint of (\ref{tracelesspi}) 
also vanishes, because $g_{ij}a^{ij}=i\pi_i^i$. 
\item $a^{ij}$, $Q^I$ and $\CH_i$ form a closed algebra of first-class 
constraints. First, $a^{ij}$ and $Q^I$ all commute. Moreover, their commutator 
with $\CH_i$ simply states how $a^{ij}$ and $Q^I$ transform under spatial 
diffeomorphisms, and therefore vanishes when the constraints are satisfied.  
\end{itemize}

Quantum mechanically, the physical wavefunctions of the membrane states should 
be annihilated by all the constraints.  Our intended ground-state wavefunction 
\be
\Psi_0[g_{ij}(\bx),X^I(\bx)]=\exp\left\{-\frac{1}{8\pi\alpha'}\int d^2\bx\,
\sqrt{g}g^{ij}\p_iX^I\p_jX^I\right\}
\ee
satisfies the quantum version of the constraint equations,
\bea
a^{ij}\Psi_0&\equiv&\left(\frac{\delta}{\delta g_{ij}}
-\frac{\sqrt{g}}{8\pi\alpha'}T^{ij}\right)\Psi_0=0,\cr
Q^I\Psi_0&\equiv&\left(\frac{\delta}{\delta X^I}
-\frac{\sqrt{g}}{4\pi\alpha'}\Delta X^I\right)\Psi_0=0,
\eea
as well as $\CH_i\Psi_0=0$.  It appears to be the only normalizable 
wavefunction satisfying all the constraints.  As a result, the spectrum of 
membrane states will contain only the ground state, and no physical excited 
states.  

This indicates that the quantization with this strong set of 
constraints provides an affirmative answer to the original question, about the 
existence of a membrane theory whose ground-state wavefunction on a Riemann 
surface $\Sigma_h$ reproduces the partition function of the bosonic string on 
$\Sigma_h$.  

\subsection{Generalizations}

The set of constraints which we imposed in the previous section is almost 
certainly unnecessarily strong.  However, it does lead to the desired result, 
a membrane theory which reproduces the string partition functions.  The  
resulting membrane theory therefore represents a solution of the 
first-quantized version of the auxililary problem posed in the introduction, 
albeit perhaps not the most exciting one:  The only physical excitation of the 
membranes are their ground states.  

Here we present a few preliminary remarks 
which might be useful in trying to find more interesting realizations, with 
physical membrane states beyond the ground state.  

In the theory without Weyl invariance, studied in Section~\ref{seclifgrav}, 
it was natural to treat $N$ as a function of only $t$, which resulted in a 
simple algebra of constraints.  Since Weyl transformations act on $N$, in a 
theory with Weyl symmetry $N$ must be allowed to depend on $t$ and $\bx$.  
However,  
\be 
\tilde N=\frac{N}{\sqrt{g}}
\ee
is an invariant under the Weyl transformations, and we may attempt to 
restrict $\tilde N$ to be a function of only $t$.  Such a restriction would 
not be fully invariant under all diffeomorphisms, but only under those 
that satisfy
\be
\p_i\zeta^i(t,\bx)=0.
\ee
These are the area-preserving diffeomorphisms of $\Sigma$.  Under this 
restriction, the algebra of Hamiltonian constraints again closes on 
$\CH_i(t,\bx)$ and $\CH_0$.  

This scenario is closely related to the possibility of not imposing Weyl 
invariance.  This in turn implies that we can move away from the conformal 
value of the coupling, $\lambda=1/2$.  As we saw in Section~\ref{secspec}, 
this set of gauge symmetries leads to one additional degree of freedom, the 
conformal factor $\phi$ of the metric.  In conformal gauge, we can write 
$g_{ij}=e^\phi\delta_{ij}$.  In string theory, $\phi$ is known as the 
Liouville field.  The kinetic term for this extra Liouville scalar $\phi$ 
will be of second order in time derivatives, $\sim (\dot\phi)^2$.  The 
analysis of Section~\ref{secspec} shows that as we move away from the 
conformal value $\lambda=1/2$ to larger $\lambda$, the sign of the kinetic 
term for $\phi$ is negative.  Thus, in this range of $\lambda$, $\phi$ plays 
the role of an extra target dimension, a phenomenon reminiscent of the 
behavior of the 
Liouville mode in noncritical string theory.  The vanishing of the Hamiltonian 
constraint $\CH_0$ on physical states will then impose an on-shell condition, 
which can be solved by membrane states with non-zero frequencies and non-zero 
spatial momenta in the target spacetime with coordinates $(X^I,\phi)$.  
In addition, if the worldsheet theory described by the classical Polyakov 
action $W$ has a conformal anomaly (as in the case of noncritical string 
theory), the effective action in two dimensions contains a nonlocal term
\be
W'\sim \int d^2\bx\,R\frac{1}{\Delta}R.
\ee
In conformal gauge, $\delta W'/\delta \phi\sim \Delta\phi$.  As a result, 
when we apply the logic of our construction to $W+W'$, the anomalous term $W'$ 
will give rise to a nonzero contribution $\sim(\Delta\phi)^2$ in the potential 
term $S_V$ of the $2+1$ dimensional action.  In conformal gauge, the Liouville 
conformal factor thus becomes a full-fledged Lifshitz scalar.  

Note that the effective metric on the target manifold $(X^I,\phi)$ will be 
relativistic, as can be seen from the worldvolume kinetic terms for these 
fields, schematically of the form $\dot X^2-\dot\phi^2$.       
If such a theory can be consistently quantized, it is likely to produce 
a relativistic spectrum of low-frequency modes, for which we would have 
a natural interpretation, as the superfuid excititations of the 
second-quantized Bose-Einstein condensate discussed in Section~2.      

The full quantum theory of membranes in the noncritical regime, with the 
Liouville field $\phi$ as one of the dynamical degrees of freedom and playing 
the role of time, is likely to be very difficult to analyze, with 
complications similar to those that occur in string theory away from its 
critical dimension.  

\section{Conclusions}

In this paper, we have introduced a new class of nonrelativistic gravity 
models, characterized by anisotropic scaling between space and time with a 
nontrivial value of the dynamical critical exponent $z=2$.  This anisotropy 
leads to a change of the critical dimension of the system to $2+1$, and makes 
the theory suitable for the worldvolume of a membrane where it can be coupled 
to quantum critical matter with $z=2$.  

Any mathematically consistent theory of gravity can be expected to have 
at least four different categories of applications: 

\begin{itemize}
\item[(i)] On worldvolumes of strings or branes, as required by 
their worldvolume reparametrization invariance.
\item[(ii)] As a theory of the observed gravitational effects in our 
Universe.  
\item[(iii)] In the context of the AdS/CFT correspondence, as a candidate for 
the dual description of interesting classes of CFTs and more general quantum 
field theories.  
\item[(iv)] Applications in mathematics, such as those produced by topological 
gravity and topological strings.  
\end{itemize}
The present paper mostly focused on the first class of applications of 
nonrelativistic gravity, as a candidate theory on the membrane worldvolume 
where the $z=2$ system is at its critical dimension.  However, our more 
general discussion of gravity at $z=2$ in $D+1$ spacetime dimensions in 
Section~\ref{seclifgrav} can be expected to be useful for possible 
applications (ii) and (iii) as well (see also \cite{lif}).  

As to (iv), we have seen in Section~\ref{secdetbal} that gravity at the $z=2$ 
Lifshitz point is intimately related to the Ricci flow equations, and in a 
sense represents the natural quantum field theory associated with the Ricci 
flow.%
\footnote{Another relation between Perelman's theory and quantum field 
theory was explored by Tseytlin \cite{tseytlin}.}
The concept of the Ricci flow was instrumental in Perelman's theory 
and the proof of the Poincar\'e conjecture \cite{perelman}.  It would be 
interesting to develop this connection further, and see for example whether 
correlation functions of natural observables in our field theory shed 
additional light on Perelman's theory.  Our theories of gravity with 
anisotropic scaling should also be relevant to the mathematically rich theory 
of foliations and their invariants \cite{lawson,godbillon,moerdijk}.  

In the context of $z=2$ worldvolume gravity, the problem of summing over 
membrane topologies and organizing the sum into a topological expansion is 
also put in a new light:  The 3-manifolds in question now carry an additional 
topological structure of a foliation.  It is possible that this extra 
structure makes the summation over a specific class of foliated manifolds 
more managable that the sum over all topologies.  When membranes interact, the 
topology of the spatial leaves of the foliation changes.  Hence, in the sum 
over topologies, membrane interactions are likely to require foliations with 
singularities, such as those that occur in Morse theory \cite{morse}, with the 
role of the Morse function played by worldvolume time.  

In this paper, we only considered the simplest case, of the bosonic theory.   
In order to see whether the ideas of anisotropic worldvolume gravity are 
relevant to the relativistic M2-branes of supersymmetric M-theory, a 
generalization of our framework to membranes with spacetime supersymmetry 
would be required.  In particular, it is natural to ask whether any version 
of anisotropic gravity in $2+1$ dimensions can flow naturally to $z=1$ at long 
distances and serve as a UV completion of the relativistic worldvolume theory 
\cite{memb} on the membranes of M-theory.  

\acknowledgments
Results reported in this paper were first presented in a talk delivered 
on November~11,~2006 at the {\it M-Theory in the City\/} workshop at the 
University of London, organized to mark the 11th anniversary of the 
discovery of M-theory; at the Sowers Workshop on {\it What is String 
Theory?\/} at Virginia Tech (May 2007); at the Banff workshop on {\it New 
Dimensions in String Theory\/} (June 2008); at the {\it AdS, Condensed Matter 
and QCD\/} workshop at McGill (October 2008); and in talks at LBNL (December 
2006), KITP (November 2007), Stanford (December 2007), Masaryk University 
(July 2008) and MIT (October 2008).  I wish to thank the organizers for their 
hospitality, and the participants for stimulating discussions.  At various 
stages of this work, I benefitted from discussions with Jan de~Boer, 
M\aa ns Henningson, Charles Melby-Thompson, Kelly Stelle, and Edward Witten.  
This work has been supported by NSF Grant PHY-0555662, DOE Grant 
DE-AC03-76SF00098, and the Berkeley Center for Theoretical Physics.  

\bibliographystyle{JHEP}
\bibliography{mqc}
\end{document}